\documentclass[aps,prb,groupedaddress, twocolumn, notitlepage, portrait]{revtex4-2}
\usepackage[utf8]{inputenc}
\usepackage[T1]{fontenc}
\usepackage{amsmath}
\usepackage{amsfonts}
\usepackage{amssymb}
\usepackage{braket}
\usepackage{graphicx}
\usepackage{tabularx}
\usepackage{enumitem}
\usepackage[colorlinks=true, citecolor=blue, linkcolor=blue, urlcolor=blue]{hyperref}
\begin{document}
\title{Laughlin-like states of few atomic excitations in small subwavelength atom arrays}
\author{B\l a\.{z}ej Jaworowski}
\email{blazej.jaworowski@icfo.eu}
\affiliation{ICFO - Institut de Ciencies Fotoniques, The Barcelona Institute of Science and Technology, Castelldefels (Barcelona) 08860, Spain}
\author{Darrick E. Chang}
\affiliation{ICFO - Institut de Ciencies Fotoniques, The Barcelona Institute of Science and Technology, Castelldefels (Barcelona) 08860, Spain}
\date{\today}

\begin{abstract}
 Atom arrays with sub-wavelength lattice constant can exhibit fascinating optical properties. For example, the combination of $V$-type level structure and magnetic fields can yield topological band structures, making the neutral atomic excitations behave like charged particles in a magnetic field. Up to now, much of our understanding of these systems (and arrays in general) focuses on the single-excitation regime. Here, we go beyond the single-excitation level to show that such systems can give rise to few-particle Laughlin-like states. In particular, we consider small honeycomb ``flakes,'' where the harmful divergences near the light cone can be smeared out by finite-size effects. By choosing an appropriate value of magnetic field we thereby obtain an energy spectrum and eigenstates resembling those of Landau levels. The native hard-core nature of atomic excitations then gives rise to multi-excitation Laughlin-like states. This phenomenon occurs not only in samples of tens of sites, but also in a minimal nanoring system of only six sites. Next, considering two-particle Laughlin-like states, we show that they can be driven by uniform light, and that correlations of the output light contain identifying fingerprints of these states. We believe that these results are a step towards new paradigms of engineering and understanding strongly-correlated many-body states in atom-light interfaces.
\end{abstract}

\maketitle

\section{Introduction}

Atom arrays with sub-wavelength lattice constant are a new frontier in quantum optics. At such spacings, the atoms start to influence each other via long-range dipole-dipole interaction, and spontaneous emission becomes a collective phenomenon. This leads to effects such as dipole blockade \cite{cidrim2020photon,williamson2020superatom} or perfect reflection from a single atom layer \cite{rui2020subradiant}. Moreover, it was proposed that such collective phenomena can be harnessed to drastically improve the efficiency of quantum memories \cite{asenjo2017exponential,manzoni2018optimization} and photon gates \cite{moreno2021quantum}. In the presence of many excitations, nonlinearities arise due to the hard-core nature of the atomic level structure, as a given atom can only be excited once. Despite active efforts to tackle the many-excitation case~\cite{clemens2003collective,ostermann2012cascaded, bettles2020quantum,qu2019spin,zhang2019theory,mahmoodian2020dynamics,masson2020many, masson2022universality,rubiesbigoda2023dynamic,holzinger2025symmetry}, this is still, to a large extent, an unknown territory. Fundamental questions, such as what kinds of many-body effects can be exhibited by sub-wavelength arrays, and what kinds of theoretical methods can be used to describe them, remain open for exploration.

In the search for new approaches, one can look for inspiration in condensed matter physics and the considerable knowledge built there. One particular instance of a strongly correlated paradigm that has generated substantial interest is the physics of topological orders \cite{wen1990topological}, for example the fractional quantum Hall (FQH) states \cite{tsui1982two,laughlin1983anomalous} that occur in two-dimensional electron gases at high magnetic field. Such states, lacking order in the conventional Landau sense, display many nonintuitive physical phenomena, such as excitations with fractional charge and fractional statistics \cite{laughlin1983anomalous,arovas2984fractional}. The latter is not only of theoretical interest, but is thought to serve as a basis for the concept of topological quantum computation, a hypothetical way of making quantum computers fault-tolerant \cite{kitaev2003fault}. 

Can we find an analogy of topological order in a quantum optical array? Such an example might be beneficial for both quantum optics and condensed matter. In particular, it can provide new insights into how to generate strongly correlated effects using light, and can also provide new opportunities to realize and probe FQH states experimentally. We note that there are already ongoing attempts to realize such states~(so far successful at the level of two particles) in other quantum simulation platforms~\cite{clark2020observation,leonard2023realization, lunt2024realization,wang2024realization}.

In this work, we discuss a way to realize Laughlin-like states in sub-wavelength arrays. Beyond a 2D electron gas, the FQH effect can occur starting from single-particle energy bands that mimic a Landau level \cite{sheng2011fractional, neupert2011fractional}, in particular if the bands are flat and have nonzero Chern number. It is already known that topological bands can be created by applying magnetic fields to arrays of three-level atoms \cite{perczel2017topological,perczel2017photonic,bettles2017topological}. We show that they can be made flat, except near the light cone, and that the divergence at the light cone can be smeared by finite-size effects in small systems. Combined with the native hard-core interaction, this gives rise to two- and three-particle fractional quantum Hall states, which we confirm numerically using exact diagonalization. Inspired by theoretical~\cite{repellin2020fractional} and experimental~\cite{clark2020observation,leonard2023realization, lunt2024realization,wang2024realization} work in other systems showing that even in few-particle systems one can observe some characteristic properties of the FQH states, we show that the states found in small arrays display high overlap with model states and level counting characteristic to fractional exclusion statistics. Surprisingly, a system as small as only six sites is enough to obtain these characteristic features. 

Next, inspired by work on FQH states of photons in optical cavities \cite{umunculiar2014probing, umuncalilar2017generation, clark2020observation}, we show that the output light from our systems carries information about the angular momenta of the underlying single-particle states and two-particle Fock states, which is enough to differentiate these states from other eigenstates. The states are also accessible by driving, and the output light from the steady state shows characteristic features which allow to identify the two-excitation component as being close to the Laughlin-like ground state. 

\section{Single-particle picture}\label{sec:singleparticle}

\subsection{The system}

\begin{figure}
    \centering
    \includegraphics[width=\linewidth]{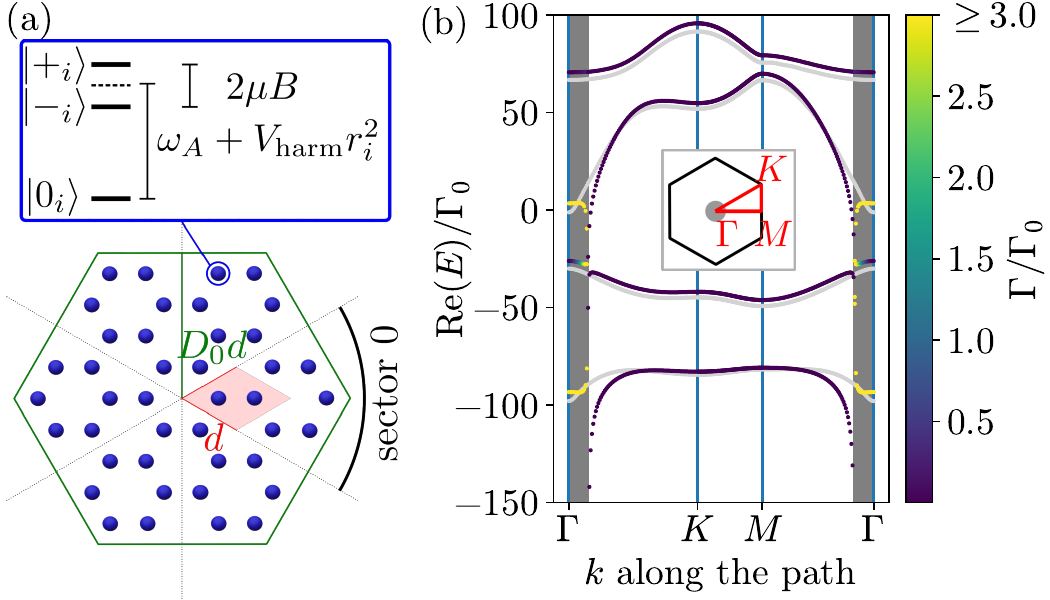}
    \caption{(a) A flake of a honeycomb array of three-level atoms. In the blue frame, the energy levels of a single atom are shown. The unit cell of the hexagonal lattice (a rhombus with edge length $d$) is shown in light red. Green lines show the edges of the flake (a hexagon with inscribed circle of radius $D_0d$). The ``sector 0'' is a starting piece of the flake from which we can generate the whole flake by successive rotations by $\pi/3$.
 (b) The band structures of the simplified model (only $\sim 1/r^3$terms retained in Eq.~\eqref{eq:GreensFunction}; light gray) and the full model (full Eq.~\eqref{eq:GreensFunction}; other colors) for dimensionless lattice constant $d_0=0.1$ (other colors) at rescaled magnetic field $B_0=12$. In the latter case, the colors denote the decay rate due to spontaneous emission. The gray region denotes the light cone, i.e. the part of the Brillouin zone such that $|\mathbf{k}|<k_A$. The vertical blue lines denote high-symmetry points of the Brillouin zone (see the inset).}
    \label{fig:SystemPicture}
\end{figure}

We consider finite flakes of an array of three-level atoms arranged in a honeycomb lattice and subject to a perpendicular magnetic field, as shown in Fig.~\ref{fig:SystemPicture}(a).  We consider a simple model of an atom with a unique electronic ground state and exhibiting an isotropic response to light, as captured by three excited states. We choose to represent two of these states in atom $i$ by $\ket{\pm_i}=a_{i\pm}^{\dagger}\ket{0_i}$, where $\ket{+_i}$ represents the state that can be optically excited by circularly polarized ($\sigma^+$) light (analogously $\sigma^-$ for $\ket{-_i}$). In the absence of a magnetic field, these states are degenerate at energy $\omega_A$ relative to the ground state $\ket{0_i}$ (we set $\hbar=1$ throughout this work). In the presence of a magnetic field $B$, the degeneracy is lifted and the energies of states $\ket{\pm_i}$ are $\omega_A \pm \mu B$, where $\mu$ is the magnetic moment of the atoms. The third excited state will couple to light polarized out of the plane, and can thus be ignored since we will only consider driving fields that are circularly polarized in-plane.

If the atoms are sufficiently close together, an excited atom can emit a photon that in turn mediates an interaction with another atom. To describe these processes, we will consider an array with a lattice constant $d$ associated with the unit cell~(see Fig.~\ref{fig:SystemPicture}(a)). We also define a dimensionless lattice constant $d_0=d/\lambda_A$, where $d_0<1$ and $\lambda_A=\frac{2\pi c}{\omega_A}$ is the wavelength of the atomic transition (note that the nearest-neighbor interatomic distance is $d/\sqrt{3}$). Without loss of generality, in the numerical calculations we will use $\lambda_A=1$, in which case the associated wavenumber is $k_A=2\pi$.

Within the ``spin-model'' formalism \cite{dung2002resonant, xu2015input, caneva2015quantum, asenjo2017exponential}, where the photons are integrated out within the Born-Markov approximation where retardation is ignored, one can write down an effective master equation describing the evolution of the atom-only density matrix $\hat{\rho}$,
\begin{equation}
    \dot{\hat{\rho}}=-i(\hat{H}\hat{\rho}-\hat{\rho} \hat{H}^{\dagger})+2\sum_{ij}\Gamma_{ij\sigma_1\sigma_2}\hat{a}_{j\sigma_2} \hat{\rho} \hat{a}_{i\sigma_1}^{\dagger}
    \label{eq:master}
\end{equation}
where the first two terms describe an evolution according to an effective non-Hermitian Hamiltonian $\hat{H}$, while the third term is a population-recycling term depending on collective decay rates $\Gamma_{ij\sigma_1 \sigma_2}$~(which need not be specified for our purposes). Physically, the non-Hermitian nature of $\hat{H}$ and the population recycling terms reflect that coupling with light can give rise to correlated dissipation or spontaneous emission processes, associated with interference in emission between different atoms. 

In our case, the effective non-Hermitian Hamiltonian has the form 
\begin{equation}
    \hat{H}=\hat{H}_\mathrm{band}+\hat{H}_\mathrm{harm}      
    \label{eq:ham}
\end{equation}
where $\hat{H}_\mathrm{band}$ is the translationally invariant part of the Hamiltonian, and $\hat{H}_\mathrm{harm}$ is an effective harmonic confining potential. The former is given by
\begin{multline}
    \hat{H}_\mathrm{band}=-i\Gamma_0/2\sum_i \hat{n}_i+ \mu B\sum_i(\hat{n}_{i+}-\hat{n}_{i-})\\-\frac{3\pi \Gamma_0}{k_A} \sum_{i\neq j}\sum_{\sigma, \sigma'}
G_{\sigma, \sigma'}(\mathbf{r}_{i}-\mathbf{r}_{j})
a^{\dagger}_{i\sigma}a_{j\sigma'},
    \label{eq:hband}
\end{multline}
where $\hat{n}_{i\sigma}=\hat{a}^{\dagger}_{i\sigma}\hat{a}_{i\sigma}$ is the occupation of a given atomic orbital on a given site, $\hat{n}_{i}=\hat{n}_{i+}+\hat{n}_{i-}$ is the total site occupation, $B$ is the magnetic field, and $G_{\sigma, \sigma'}(\mathbf{r})$ is the dyadic electromagnetic Green's function (see \eqref{eq:GreensFunction} below) and $\Gamma_0=\frac{\mathcal{D}^2k_A^3}{3\pi \epsilon_0}$ is the spontaneous emission rate, with $\mathcal{D}$ being the dipole moment of the transition (we assume that the transition to both $\ket{-_i}$ and $\ket{+_i}$ states is characterized by the same dipole moment) and $\epsilon_0$ being the vacuum electric permittvity. The first term in \eqref{eq:hband} describes single-atom spontaneous emission, while the second term describes the splitting of atomic energy levels due to the magnetic field. The third describes the effect of photon-mediated interactions, both coherent and dissipative. Naturally, the strength of the interaction between two atoms $i,j$ is proportional to the electromagnetic Green's function characterizing how light propagates from one atom to the other. In the circular basis, one has 
\begin{align}
& G_{++}(\mathbf{r})=G_{--}(\mathbf{r})=\frac{e^{ik_Ar}}{8\pi r}\left(1-\frac{i}{k_Ar}+\frac{1}{(k_Ar)^2}\right),\nonumber \\ 
& G_{-+}(\mathbf{r})=\frac{e^{ik_Ar}}{8\pi r}\left(-1-\frac{3i}{k_Ar}+\frac{3}{(k_Ar)^2} \right)\exp(-2i\phi),\nonumber \\
& G_{+-}(\mathbf{r})=\frac{e^{ik_Ar}}{8\pi r}\left(-1-\frac{3i}{k_Ar}+\frac{3}{(k_Ar)^2} \right)\exp(2i\phi).
\label{eq:GreensFunction}
\end{align}
where $r$ and $\phi$ are the components of vector $\mathbf{r}$ in polar coordinates.

A harmonic potential is added to confine atomic excitations near the center and thereby make the edges of the system smooth,
\begin{equation}
    \hat{H}_\mathrm{harm}=V\sum_{i}r_i^2 \hat{n}_{i}
\end{equation}
with $r_j$ being the distance between the atom and the center of the system. In practice, this could be implemented as an ac Stark shift associated with a far-detuned laser. 

The most general dynamics would require the density matrix of Eq.~\eqref{eq:master} to be solved. Here, however, we will only be interested in weak driving, where most of the atomic population remains in the ground state $|0_i\rangle$, with a small probability to find one atom in an excited state and even smaller to find two atoms in an excited state. In that case, it is well-known that one can avoid solving the full master equation, and obtain information about the steady state purely by considering the non-Hermitian Hamiltonian \cite{carmichael1991quantum,hafezi2013nonequlilbrium}.

\subsection{Band structure of an inifinite system}

It is instructive to first look at the case of a single particle (i.e. a single atomic excitation) in an infinite system described by $H_\mathrm{band}$ only. It has previously been shown that this system can host topologically nontrivial energy bands \cite{perczel2017topological,perczel2017photonic,bettles2017topological}. Here we want to consider whether these bands can be made flat. Let us consider a simplified model, which assumes small lattice constant $d_0\ll 1$. Let us suppose that the $\sim 1/r^3$ terms of the Green's function dominate, and $e^{ik_Ar}\approx 1$. Such a model was studied in Ref.~\cite{weber2022experimentally} in the context of Rydberg atoms. It was shown that at certain values of magnetic field, this model has a nearly-flat, topological lowest band which can host a Laughlin FQH state (see the light gray lines in Fig.~\ref{fig:SystemPicture}(b)). If we define a rescaled magnetic field $B_0=\frac{d^3k_A^3\mu}{\Gamma_0} B$, the energy spectrum at fixed $B_0$ becomes independent of the interatomic distance, up to multiplication by a constant. The lowest band is topological (Chern number $C=1$) and nearly flat at $B_0\approx 12$. We will use $B_0=12$ in subsequent calculations.

We note that such a topological flat band is a characteristic feature of the honeycomb lattice. We studied the $d_0\ll 1$ limit also in triangular and square lattice and did not find a situation where a comparable topological flat band could be generated, unless a strong onsite potential is applied to one third of sites of the triangular lattice, which likens the model to the honeycomb lattice.

If, instead, we consider the full model with Green's function \eqref{eq:GreensFunction}, and plot the real part of the eigenvalues of $H_\mathrm{band}$ (see the colorful lines in Fig.~\ref{fig:SystemPicture}(b)), one can see a qualitative difference with respect to the simplified model: the lowest band diverges around $|\mathbf{k}|=k_A$. In the exact description of our system the eigenstates are polaritons being linear combinations of atomic and photonic modes. In particular, one branch of polaritons is light-like at $|\mathbf{k}|\approx 0$, i.e. for low $|\mathbf{k}|$ they are composed mostly of photons and have light-cone dispersion $\omega=ck$. This branch anticrosses with other polariton branches near $|\mathbf{k}|=k_A$. In our approximation that retardation effects associated with light can be ignored, the dispersion relation $\omega=ck$ becomes a vertical line at $|\mathbf{k}|=k_A$, and the anticrossing becomes a divergence. Also, by tracing out the photons, the light-like polaritons disappeared from the picture, leaving only the atom-like ones.

While the location $|\mathbf{k}|=k_A$ of the light cone remains constant, the size of Brillouin zone is inversely proportional to $d_0$. Therefore, the relative size of the region within the light cone, $|\mathbf{k}|<k_A$, compared to the Brillouin zone shrinks as we decrease $d_0$. With small $d_0$ (such as $d_0=0.1$ in Fig.~\ref{fig:SystemPicture}(b)), the band structure outside the light cone is similar to the one of the simplified model. In the limit $d_0\rightarrow 0$, the simplified model is recovered. Therefore, to reduce the influence of the light cone, we should consider as small $d_0$ as possible. 

Besides creating a divergence for the lowest band, the light cone also has the important effect of causing Bloch states within the light cone to couple to radiation fields and therefore experience a non-zero decay rate. Here, we define the decay rate as $\Gamma=-2\mathrm{Im} E$, where $E$ is the eigenvalue of the non-Hermitian Hamiltonian $H_\mathrm{band}$. We indicate the decay rate of Bloch states by their color in Fig.~\ref{fig:SystemPicture}(b). In general, the values of $\Gamma$ within the light cone far exceed the spontaneous emission rate $\Gamma_0$.

\subsection{Finite flakes}

To suppress the effect of the divergence, we consider finite, hexagonal systems of size $D_0d$, as shown in Fig.~\ref{fig:SystemPicture}(a). The rationale is that if the system is small enough, the long-range terms which cause the divergence cannot fully manifest themselves and thus the finite-size effects smear the degeneracy.

\begin{figure}
    \centering
    \includegraphics[width=\linewidth]{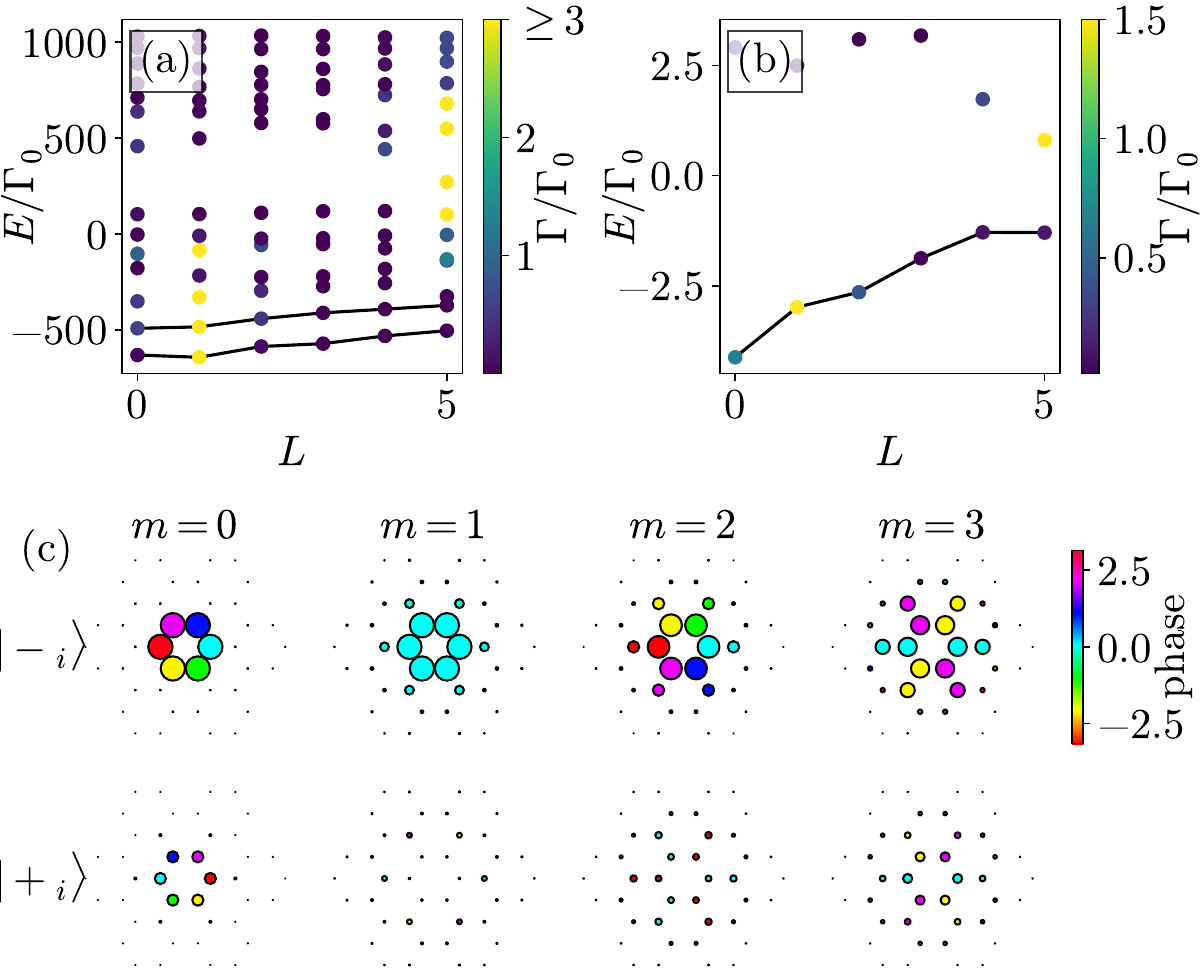}
    \caption{(a) Spectrum of the Hermitian part of the Hamiltonian of a $D_0=2.1$ flake with $V_\mathrm{harm0}=3$ at $d_0=0.05$. (b) Spectrum of the Hermitian part of the Hamiltonian of a $D_0=0.6$ flake at $d_0=3$. The black lines in (a) and (b) are a guide to the eye, showing the energy growing with angular momentum. (c) Eigenstates corresponding to four lowest energies in (a). The columns correspond to eigenstates labeled by their angular momentum $L=m$, while rows correspond to the coefficients of the wavefunction on the two atomic orbitals. The color and area of each circle denote the phase and amplitude of the coefficient corresponding to the given site in the ``original'' basis ($\hat{a}_{i\sigma}^{\dagger}$).}
    \label{fig:FiniteSPEigenstates}
\end{figure}

We proceed with calculating the single-particle energy spectra of flakes as a function of array angular momentum $L$ (to avoid confusion with other kinds of angular momenta, see Table \ref{tab:angularmomenta}, where we differentiate between the array, continuum and light angular momenta). $L$ labels the angular momentum of the system, after a phase transformation $\hat{a}^{\dagger}_{i \sigma}\rightarrow \hat{\tilde{a}}^{\dagger}_{i \sigma}$ is applied to make Eq. \eqref{eq:GreensFunction} rotationally invariant (see Appendix \ref{app:rotsymm}). For reasons that will become clear later, we want to obtain a set of orthogonal eigenstates, and therefore we split the Hamiltonian \eqref{eq:ham} into the Hermitian and anti-Hermitian components $\hat{H}=\hat{H}_{\mathrm{herm}}+\hat{H}_{\mathrm{anti}}$ and diagonalize the Hermitian part only. For the ease of comparing different systems, we define a rescaled harmonic potential strength as $V_\mathrm{harm0}=\frac{d^5k_A^3}{\Gamma_0} V_\mathrm{harm}$.

The resulting spectrum for $D_0=2.1$, $d_0=0.05$, $V_\mathrm{harm0}=3$ (Fig. \ref{fig:FiniteSPEigenstates}(a)) shows no sign of divergence. Instead, its low-energy part resembles the spectrum of the lowest Landau level in a harmonic potential, with energy growing with array angular momentum (defined modulo 6, so the low-energy ``branch'' reaches $L=5$ and then continues from $L=0$), see the black lines in the plot.  The low-energy wavefunctions (Fig. \ref{fig:FiniteSPEigenstates}(c)) resemble the angular momentum orbitals of the lowest Landau level (LLL) in the symmetric gauge \cite{chakraborty1995quantum}, in the sense that they are ring-shaped and their radius grows with angular momentum. One can write the correspondence in the following way \cite{he2015wavefunction}, 
\begin{equation}
    \ket{\phi^i_L} \leftrightarrow \ket{\phi^{LLL}_{m=6i+L}}
    \label{eq:correspondence}
\end{equation}
where we denoted the $i$th single particle eigenstate ($i=0,1, \dots$) of the array at array angular momentum $L$ as $\ket{\phi^i_L}$, while $\ket{\phi^{LLL}_{m}}$ is the symmetric-gauge lowest Landau level orbital with continuum angular momentum $m$ (described in detail in Sec. \ref{sec:laughlinlike}). Thus, we will call single-particle eigenstates of the array ``orbitals'' (and orbitals of single atoms will be referred to as ``atomic orbitals''). Such correspondence was found before in the topological flat band models \cite{he2015wavefunction}.

To understand the possible response of the different orbitals to light, it is convenient to return to the ``original'' basis defined by operators $\hat{a}_{i\sigma}^{\dagger}$ rather than the transformed $\hat{\tilde{a}}_{i\sigma}^{\dagger}$. In the original basis, the $L=1$ orbital has a constant phase (at least on the$\ket{-_i}$ atomic orbital where the majority of the weight of the wave function is located). Such a lack of phase variation suggests that such a state will be strongly responsive to light. To quantify this, we define the decay rate for orbital $\ket{\phi^i_L}$ as 
\begin{equation}
\Gamma=-2\mathrm{Im} \braket{\phi^i_L
|\hat{H}_\mathrm{anti}|\phi^i_L}
\label{eq:DecayRate}
\end{equation}
 where $\hat{H}_\mathrm{anti}$ is the anti-Hermitian part of $\hat{H}$. The value of $\Gamma$ for different eigenstates is indicated by the color of the dots in Fig \ref{fig:FiniteSPEigenstates}(a). We can see that a few low-energy states at $L=1$ are indeed strongly superradiant, with a decay rate larger than that of a single excited atom, $\Gamma_0$.

Surprisingly, some properties of the 2D flakes still remain in the minimal system, $D_0=0.6$, composed of one hexagon only (Fig \ref{fig:FiniteSPEigenstates}(b)), resembling the nanoring systems previously studied in Refs \cite{asenjo2017exponential, cardoner2019subradiance}. The four-band structure is no longer visible, as such a hexagon can be understood as a periodic chain with two bands only. Obviously, the radius of eigenstates is now constant instead of growing with angular momentum. But the energy of the lower band seems to grow with angular momentum even in the absence of a confining potential (which has no effect), resembling the edge spectrum of bigger structures.

\begin{table}[]
    \centering
    \begin{ruledtabular}
    \begin{tabular}{cccccccccc}
    state  & $\color{red}\ket{\phi^0_{0}}$ & $\color{red}\ket{\phi^0_{1}}$ & $\color{red}\ket{\phi^0_{2}}$ & $\ket{\phi^0_{3}}$ & $\ket{\phi^0_{4}}$ & $\ket{\phi^0_{5}}$ & $\ket{\phi^1_{0}}$ & $\ket{\phi^1_{1}}$ & $\dots$  \\ \hline 
      $L$  &  $0$ & $1$ & $2$ & $3$ & $4$ & $5$ & $0$ & $1$ & $\dots$\\
      $L'=m$ & $0$ & $1$ & $2$ & $3$ & $4$ & $5$ & $6$ & $7$ &  $\dots$\\
      $l$ ($-$) & $-1$ & $0$ & $1$ & $2$ & $3$ & $-2$ & $-1$ & $0$  & $\dots$\\
      $l$ ($+$) &  $1$ & $2$ & $3$ & $-2$ & $-1$ & $0$ & $1$ & $2$ & $\dots$ \\
    \end{tabular}
    \end{ruledtabular}
    \caption{Comparison between different angular momenta assigned to states in this work: the array angular momentum $L$ defined modulo 6, the continuum angular momentum of a corresponding LLL orbital $L'$ (or $m$), the main component $l$ of the angular momentum of emitted/absorbed minus- and plus-polarized light (note that other components differing by multiple of 6 can also be absorbed/emitted). The three orbitals in red are the ones from which a model two-particle Laughlin state is formed. }
    \label{tab:angularmomenta}
\end{table}

\section{Laughlin-like states}\label{sec:laughlinlike}

\subsection{Model Laughlin states}\label{sec:modelstates}

Atomic excitations by definition exhibit infinite onsite repulsion: an atom either is in a ground state (zero particles), or has an excitation of the $+$ or $-$ flavor (one particle), but there is no possibility of two excitations ``occupying'' the same atom. This, together with the LLL-like states described in Sec.~\ref{sec:singleparticle}, raises hopes that in the case of $N_\mathrm{part}>1$ excitations, Laughlin-like states will form. Indeed, in Ref.~\cite{weber2022experimentally} the emergence of the Laughlin states was observed in the Rydberg model corresponding to the $d_0\rightarrow 0$ limit of our systems. In this section we show that this is the case also for finite $d_0$. 

\begin{figure}
    \centering
    \includegraphics[width=\linewidth]{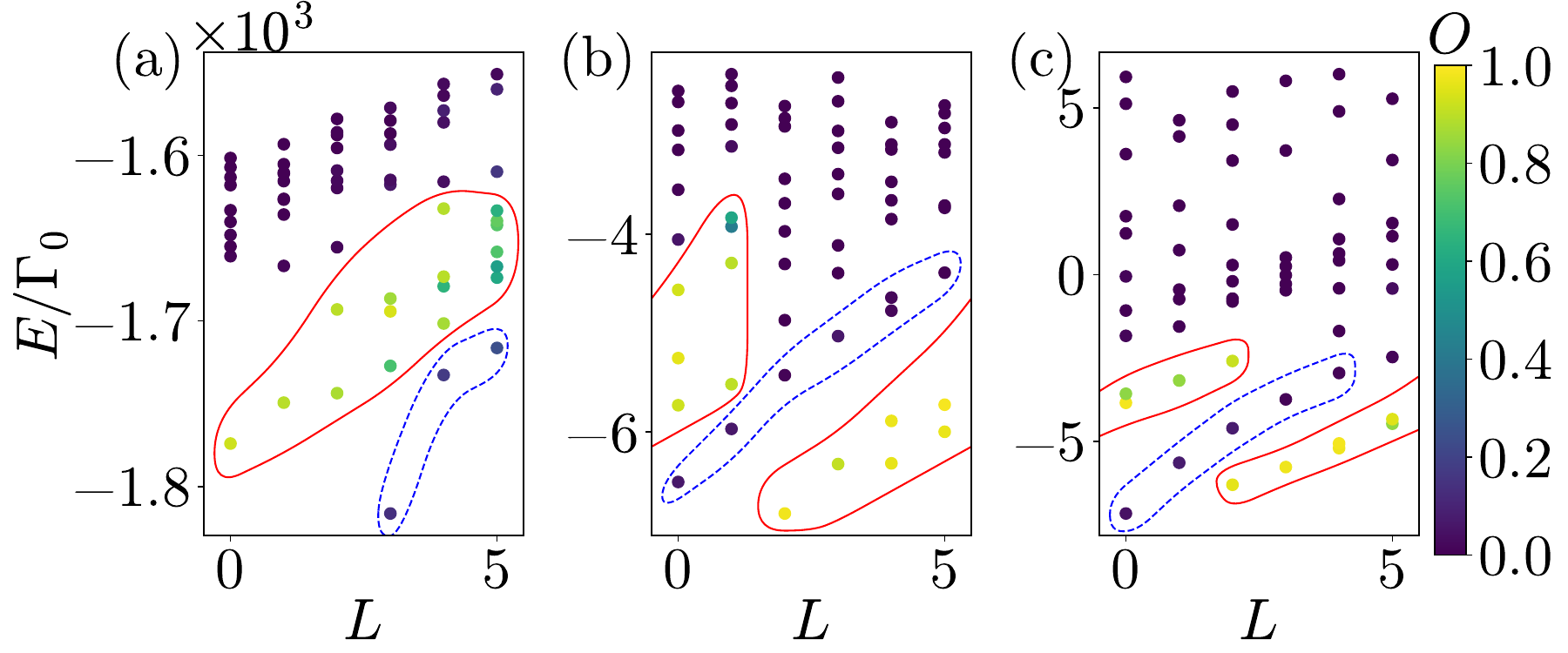}
    \caption{Energy spectra of the Hermitian part of the Hamiltonian, as a function of angular momentum, obtained for (a) $D_0=2.1$, $d_0=0.05$, $V_\mathrm{harm0}=3$, $N_\mathrm{part}=3$, (b) $D_0=1.6$, $d_0=0.3$, $V_\mathrm{harm0}=10$, $N_\mathrm{part}=2$. (c) $D_0=0.6$, $d_0=0.3$, $V_\mathrm{harm0}=0$, $N_\mathrm{part}=2$. The colors denote the sum of squared overlaps of a given ED eigenstate with model states at given angular momentum. The solid red (dashed blue) line encircles the Laughlin-like (non-Laughlin-like) branch of states. The cutoffs are: $\Delta L'_\mathrm{max}=5$ in (a), (b) and $\Delta L'_\mathrm{max}=6$ in (c), as well as $m_\mathrm{max}=9,7,5$ in (a), (b), (c), respectively.  
    }
    \label{fig:Overlap}
\end{figure}

But before we analyze the results for arrays, let us first discuss the model Laughlin states, to which later we will compare the array eigenstates. We follow the approach from Ref.~\cite{he2015wavefunction}. That is, we start from a set of states in the continuum LLL populated by $N_\mathrm{part}$ bosonic particles (analogous to atomic excitations in the array case). The $N_\mathrm{part}$-particle Laughlin ground state \cite{laughlin1983anomalous} at filling 1/2 and the edge excitations on top of it \cite{wen1990chiral} are given by,
\begin{equation}
    \Psi=P(\mathbf{z})\prod_{i}^{N_\mathrm{part}}\prod_{j=i+1}^{N_\mathrm{part}}(z_i-z_j)^2\exp\left(- \frac{\sum_j|z_j|^2}{4l_B^2}\right)
    \label{eq:laughlin}
\end{equation}
where $z_i=x_i+iy_i$ is the position of the $i$th particle written as a complex number, $l_B$ is the magnetic length and $P(\mathbf{z})$ is an arbitrary symmetric polynomial in the $z$ variables. For the ground state, $P(\mathbf{z})=1$. For edge excitations with fixed continuum angular momentum $L'=L_0'+\Delta L'$, where  $L_0'$ is the continuum angular momentum of the ground state, $P(\mathbf{z})$ is a homogenous polynomial of degree $\Delta L'$. The fact that typically at given $\Delta L'$ there are a number of independent polynomials reflects the degeneracy of the edge modes \cite{wen1990chiral}, which in turn is related to the fractional exclusion statistics of the quasiholes (anyonic excitations of the Laughlin state) \cite{haldane1991fractional}. 

To be able to relate to an atom array, we now convert Eq.~\eqref{eq:laughlin} to the second-quantized description in the basis of Fock states composed of the lowest Landau level angular momentum orbitals, $\phi^{LLL}_{m}(z)=\frac{1}{\sqrt{2\pi l_B^2 2^mm!}}\left(\frac{z}{l_B}\right)^m \exp\left(- \frac{|z|^2}{4l_B^2}\right)$.  Since these orbitals depend only on $z^m$ (except for normalization and common factors), Eq. \eqref{eq:laughlin} can be cast into second quantized form by expanding the polynomial, identifying symmetric combinations of monomials as Fock states and taking care of normalization. We note that the power $m$ is the orbital's continuum angular momentum, therefore the total continuum angular momentum of state \eqref{eq:laughlin} can be obtained by expanding the polynomial and adding the powers $m_i$, $L'=\sum_i m_i$, in any of the terms. Then, we replace the orbitals by the single-particle eigenfunctions of the array (Fig.~\ref{fig:FiniteSPEigenstates}(c)) according to Eq. \eqref{eq:correspondence}, resulting in a model Laughlin wavefunction $\ket{\Psi_{\mathrm{model}}}$, mapped to the array system. For example, for a two-particle Laughlin ground state we get
\begin{equation}
  \ket{\Psi_{\mathrm{model}}}=\frac{1}{\sqrt{2}}\left(\ket{\phi^0_0 \phi^0_2}-\ket{\phi^0_1 \phi^0_1}\right)  
  \label{eq:laughlin_2part}
\end{equation}

We note that in general there is a phase ambiguity in \eqref{eq:correspondence}, which can influence the value of the overlap (e.g. after multiplying $\ket{\phi^0_1}$ by $i$, \eqref{eq:laughlin_2part} maps to a $\sim (z_1+z_2)^2$ wavefunction instead of $\sim (z_1-z_2)^2$). To resolve it, we choose the following phase convention: just like the Landau level wavefunctions  $\phi^{LLL}_{m}(z)$ have zero phase on the positive real axis, we choose that the array single-particle wavefunction has zero phase on the $\ket{-_i}$ atomic orbitals in the ``sector 0'' (see Fig. \ref{fig:SystemPicture} (a)).

In the context of the minimal 6-site system, which can be understood as a 1D chain of length 6, we also consider a 1D version of the Laughlin state arising in the ``quantum Hall circle'' limit \cite{bergholtz2009quantum}, 
\begin{equation}
    \Psi=P(\mathbf{z})\prod_{i<j}(z_i-z_j)^2
    \label{eq:laughlin1D}
\end{equation}
where the positions $z_i$ are confined to the circle $|z_i|=1$. Then, we proceed with nearly the same procedure as above, the only exception being the LLL orbitals, which now have the form $\phi^{1D}_{m}=\frac{1}{2\pi} z^m$. The different normalization factor of these orbitals lead to modified coefficients in front of Fock states in the Laughlin states. For example, a two-particle Laughlin state now becomes
\begin{equation}
\ket{\Psi_{\mathrm{model}}}=\frac{1}{\sqrt{3}}  \left(\ket{\phi^0_0 \phi^0_2}-\sqrt{2}\ket{\phi^0_1 \phi^0_1}\right).
\label{eq:laughlin_2part_1d}
\end{equation}

\subsection{Exact diagonalization and overlaps with model states}

Next, we compute the energy spectra and eigenstates of the Hermitian part of the array Hamiltonian at fixed number of atomic excitations $N_\mathrm{part}$. Having obtained the exact eigenstates numerically, we compute the overlap between them and the model states at $N_\mathrm{part}$, obtained in Sec. \ref{sec:modelstates}. Because our scheme of detection (Sec. \ref{sec:output}) works only for two excitations, and in the driving scheme (Sec. \ref{sec:output}) the 3-particle sector is much less populated than 2-exctitation one (even less for 4 excitations etc.), we focus on small excitation numbers $N_\mathrm{part}=2,3$.

The results for three example systems are shown in Fig. \ref{fig:Overlap}. These are a $D_0=2.1$ flake with three excitations, $D_0=1.6$ flake with two excitations and $D_0=0.6$ flake, also with two excitations. The plots show energy spectra as a function of array angular momentum, with the value of the overlap represented by color. The overlap $O(\ket{\psi})$ is computed as sum of squared overlaps between the given array eigenstate and all the eligible model states at the given array angular momentum. The eligible  model states are the ones with continuum angular momenta up to $L_0'+\Delta L'_\mathrm{max}$ (with $L_0'$ being the continuum angular momentum of the Laughlin ground state), composed of $\phi^{LLL}_{m}(z)$ states up to $m_\mathrm{max}$.

In all three cases, one can observe a branch (the yellow and green dots encircled by solid red lines) of states with high overlap with Laughlin ground state and edge excitations of $N_\mathrm{part}$ bosons. The analog of the Laughlin ground state is not necessarily the true $N_\mathrm{part}$-particle ground state of the system, but it is a ground state within its $L$ subspace ($L=6\mod 6=0$ for three particles, $L=2$ for two). As in continuum FQH systems, the energy of the states within the branch grows with the continuum angular momentum of the corresponding model states, that is, it grows with $L$ until $L=5$ and then continues from $L=0$.

This ``edge branch'' is particularly stable for two particles. For $D_0=1.6$ and $D_0=0.6$ one can observe it even for lattice constants as large as $d_0=0.3$ (interatomic spacing $\sim 0.17 \lambda_0$), see Figs.~\ref{fig:Overlap}(b) and (c), respectively, although in the former case it requires quite strong harmonic potential $V_\mathrm{harm0}=10$.

We also note the presence of another branch (encircled by blue dashed lines in Fig.~\ref{fig:Overlap}), which does not contain Laughlin-like states but sometimes contains the true $N_\mathrm{part}$-excitation ground state. Inspecting it for $D_0=0.6$, we found that any particular state contains a linear combination of Fock states that would not be allowed in a continuum system, but are allowed in an array because the array angular momentum is conserved modulo 6. For example, the two-excitation $L=0$ ground state of the system from Fig.~\ref{fig:Overlap}(c) has strong contributions from the $\ket{\phi_0^0 \phi_0^0 }$ and $\ket{\phi_1^0 \phi_5^0 }$ states, which would not be possible in the continuum system, because the former has continuum angular momentum $L'=0$, while the latter has continuum angular momentum $L'=6$.

\subsection{Fractional exclusion statistics}

The counting of states in the continuum Laughlin branch reflects fractional exchange statistics of fundamental excitations (quasiholes) of the system and can be determined using the generalized Pauli principle \cite{haldane1991fractional,wen1990chiral}. That is, to determine the counting of states at a given continuum angular momentum, one should fill the single-particle orbitals with $N_\mathrm{part}$ particles in such a way that on $q$ adjacent orbitals there is at most one particle (note that this procedure determines \textit{how many} states there are, not the states themselves). Our case should match with $q=2$, with $q>2$ corresponding to other Laughlin states and $q=1$ to ordinary fermions (integer quantum Hall effect) \cite{bernevig2008model}. In the case of an infinite system, all $q$ give rise to the same counting as a function of $\Delta L'$ (which reflects the fact that all the different Laughlin states have $U(1)$ edge modes \cite{wen1990chiral}); they only differ in the starting angular momentum $L_0'=qN_\mathrm{part}(N_\mathrm{part}-1)/2$ ($q=2$ is consistent with our results). Apart from $L_0'$, the differences between countings at different values of $q$ manifest themselves in finite-size effects, i.e. in a system with finite number of orbitals \cite{hermanns2011haldane}. While all the array flakes considered by us are finite, the best system to analyze is the $D_0=0.6$ system, where the six orbitals forming the lower band are clearly separated energetically from the others (see Fig. \ref{fig:FiniteSPEigenstates}(b)). The $q=2$ generalized Pauli principle for two particles on six orbitals predicts eight states (occupation numbers 
101000, 
100100,  
100010, 010100,
100001, 010010, 
010001, 001010,
001001,
000101): two at each of continuum angular momenta $L'=4,5,6$, one at each of continuum angular momenta $L'=2,3, 7,8$. After folding modulo 6, this is precisely the counting of the states encircled by the solid red line in Fig. \ref{fig:Overlap}(c). This branch of states is well separated from all other states.

\subsection{Optical properties}

Similarly to the single-excitation states, we can determine the decay rate defined as \eqref{eq:DecayRate} for the few-excitation states. It turns out that the optical properties of the states in the Laughlin-like branch differ, depending on the orbitals from which they are composed. For example, the two-particle Laughlin-like state is strongly superradiant, because it has a strong contribution of superradiant orbital $\phi^0_{1}$. In Sec. \ref{sec:driving}, we will use this fact to drive this state. Before we do that, let us first focus on how to detect its presence.

\section{Signatures of Laughlin-like states in the output light}\label{sec:output}

\subsection{The setup}

Can one retrieve the information about the Laughlin-like nature of the array eigenstates from the output light? Based on Fig.~\ref{fig:FiniteSPEigenstates}(c), one can suspect that if we initialize the system in a single-excitation eigenstate, the output light carries the information about the array angular momentum $L$ of the state: the $-$ and $+$ polarized output light has light angular momentum $l=L-1$ and $l=L+1$, respectively. Indeed, the decomposition of output light into Laguerre-Gauss modes with well-defined angular momentum was proposed \cite{umunculiar2014probing,umuncalilar2017generation}, and used in experiments \cite{clark2020observation} as a way to confirm the existence of Laughlin states of photons in a cavity setup. 

Inspired by these works, we propose how to use angular momentum of light to identify a two-particle Laughlin-like ground state among the $L=2$ eigenstates of the system. We first focus on the minimal six-site systems which are simplest in interpretation, as there is just one lower-band state per momentum sector. Moreover, the lower- and higher-band states can be differentiated based on polarization, as they are located mostly in the $\ket{-_i}$ and $\ket{+_i}$ atomic orbitals, respectively. Also, the nanoring system with two particles exhibit Laughlin-like states at $d_0=0.3$, considerably higher than e.g. $d_0=0.05$ from Fig.~\ref{fig:Overlap} (a). The performance of our identification protocol in a bigger system, $D_0=1.6$ with 24 atoms, will be studied in Sec. \ref{ssec:bigger}.

To determine the array angular momentum of a state from the light emitted by it, we propose to detect the output photons decomposed into circularly polarized Laguerre-Gauss modes of frequency $\omega_A$, waist $w_0$ and light angular momentum $l$, which, in the $z=0$ plane, have the form
\begin{equation}
        \mathbf{E}_\mathrm{det}^{(l),\sigma}(r, \phi)|_{z=0}= E_\mathrm{LG}^{(l)}\mathbf{P}_\sigma e^{i\omega_A t}
\end{equation}
where $\mathbf{P}_\sigma$ is the polarization vector, and,
\begin{equation}
    E_\mathrm{LG}^{(l)}=E_0\left( \frac{r \sqrt{2}}{w_0} \right)^{|l|}\exp\left(-\frac{r^2}{w_0^2} \right)\exp\left( -il\phi \right), 
    \label{eq:gaussian_mode}
\end{equation}
with normalization constant $E_0=\sqrt{\frac{2}{|l|! \pi w_0^2}}$ chosen in such a way that $F_\mathrm{det}=\iint_{z=\mathrm{const}} \mathrm{d}x\mathrm{d}y  \mathbf{E}_\mathrm{det}^{*}(\mathbf{r}) \cdot \mathbf{E}_\mathrm{det}(\mathbf{r})=1$. We recall that Laguerre-Gauss modes are eigenstates of light only in the paraxial approximation. On the other hand, for $l>0$, these modes have a node at the center, and so to efficiently collect light emitted by our systems, we need small, diffraction limited beam waist $w_0\sim\lambda_A$. It is known \cite{manzoni2018optimization} that \eqref{eq:gaussian_mode} remains a good approximation for the $l=0$ at $w_0=\lambda_A$. We have extended these calculations and verified that it is a good approximation also for $l\neq 0$, see Appendix \ref{app:LG}. Therefore, in the following, we will always use $w_0=\lambda_A$.

We introduce the operator of emission into a $\sigma$-polarized mode of angular momentum $l$ \cite{manzoni2018optimization},
\begin{equation}
        \hat{E}_{l}^{\sigma}=-i\sum_i E_\mathrm{LG}^{(l)*}(\mathbf{r}_i) \hat{a}_{i \sigma}
\end{equation}

\begin{figure}
    \centering
    \includegraphics[width=\linewidth]{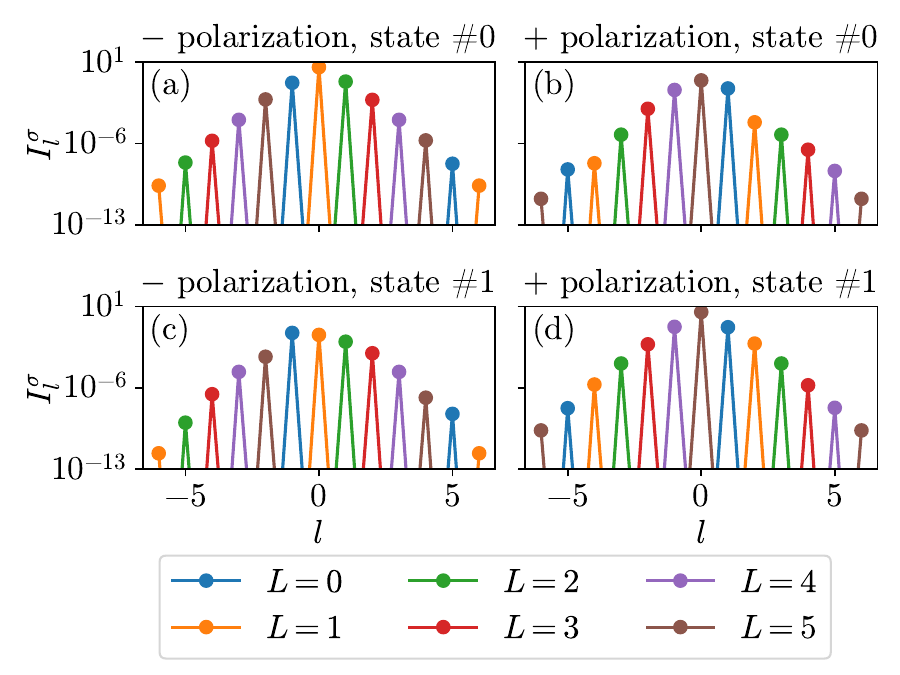}
    \caption{Intensity of light emitted from orbitals at different array angular momentum (colors) into circularly polarized Laguerre-Gauss modes with different light angular momenta $l$. Upper and lower rows correspond to the lower and upper band states, respectively, while the left and right columns show results for $-$ and $+$ circular polarization, respectively. The system parameters are $D_0=0.6$, $d_0=0.3$  and the Laguerre-Gauss modes have $w_0=\lambda_A$. }
    \label{fig:MinimalSPOutput}
\end{figure}

We denote the single-photon intensity of light emitted from the system when it is initialized in state $\ket{\psi}$ as 
\begin{equation}
    I_{l}^{\sigma}(\ket{\psi})=\braket{ \psi |\hat{E}_{l}^{\sigma\dagger}\hat{E}_{l}^{\sigma}| \psi}.
    \label{eq:I1p}
\end{equation}
For the single-excitation eigenstates (orbitals) $\ket{\phi^i_L}$, we plot the intensity $I_{l}^{\sigma}(\ket{\phi^i_L})$ in Fig.~\ref{fig:MinimalSPOutput}. One can notice that the angular momentum conservation requires that $I_{l}^{\sigma}(\ket{\phi^i_L})$ is nonzero only if $ (L-1-l)\mod 6 =0 $ ($-$ polarization) or $ (L+1-l)\mod 6 =0 $ (+ polarization). We call $f_l^{\pm}=I_{l}^{\pm}(\ket{\phi^{0}_{L=l\mp 1 \mod 6})}$, i.e. $f_l^\sigma$  is the height of the intensity peak at given $l$ in Figs.~\ref{fig:MinimalSPOutput}(a) and (b), respectively.  The intensity of the emitted light differs between states $\ket{\phi^{0}_{L}}$. In fact, $f_l^{\sigma}$ typically shrinks by orders of magnitude each time we increase $|l|$ by 1. This suppression is the stronger the smaller is the lattice constant. We assume that in our idealized setting we will be able to observe signal at least from the $|l|\leq 2$ modes.

For the two-particle states, in addition to $I_{l}^{\sigma}(\ket{\psi})$ we define two more variables: the two-photon intensity 
\begin{equation}
I_{l_1l_2}(\ket{\psi})=\braket{ \psi |\hat{E}_{l_1}^{-\dagger}\hat{E}_{l_2}^{-\dagger}\hat{E}_{l_2}^{-}\hat{E}_{l_1}^{-}| \psi}    
    \label{eq:I2p}
\end{equation}
and the two-photon amplitude 
\begin{equation}
A_{l_1l_2}(\ket{\psi})=\braket{ \emptyset |\hat{E}_{l_2}^{-}\hat{E}_{l_1}^{-}| \psi}
    \label{eq:A2p}
\end{equation}
where $\ket{\emptyset}$ is the many-atom ground state (i.e. all atoms in the atomic ground states $\ket{0_i}$). $I_{l}^{\sigma}(\ket{\psi})$ will hopefully tell us about which single-particle orbitals are occupied, $I_{l_1l_2}(\ket{\psi})$ about which two-particle Fock states (in the orbital basis) are occupied, and $A_{l_1l_2}(\ket{\psi})$ about the relative phase of the Fock state contributions. 

While $I_{l}^{\sigma}(\ket{\psi})$ and $I_{l_1l_2}(\ket{\psi})$ are fairly standard quantum optics measurements, $A_{l_1l_2}$ may be less intuitive. Nevertheless, experiments measuring similar quantities using either two-photon interference, or homodyne measurement, were proposed \cite{umuncaliar2012fractional,umunculiar2014probing}. For example, the scheme from Ref. \cite{umunculiar2014probing}, adapted to our needs, can look as follows. First, the $l=-1,0,1$ components are separated with angular momentum sorter. A phase shift $\beta$ is applied to the $l=0$ photons.  Then, $\pm 1$ holograms are applied to $\mp 1$ modes to bring them all into $l=0$. Finally, the modes are recombined using two symmetric beamsplitters and joint probability of detecting two photons on the same output arm is measured. This probability will depend on $\beta$ as a constant plus a term dependent on $\beta$, proportional to $\cos\left(2\beta+S_3\left(\ket{\psi}\right)\right)$ from which $S_3(\ket{\psi})$ can be extracted.

The modulus of $A_{l_1,l_2}$ is known from $I_{l_1,l_2}$, what remains is to determine the phase differences between $A_{l_1,l_2}$ at different $l_1,l_2$. As we show later in Sec. \ref{ssec:signatures}, the most important in our scheme is the relative phase of $A_{-1,1}$ with respect to $A_{0,0}$. Adapting the ideas of Ref. \cite{umunculiar2014probing}, one should first separate the $l=-1,0,1$ components, and then apply a phase shift of $\beta$ to the l=0 mode. Subsequently, one can apply holograms to convert the $\pm 1$ modes to $l=0$, and recombine all the modes together using two symmetric beamsplitters. The probability of detecting two photons at the same output arm will depend on $\beta$ as a constant plus a term dependent on $\beta$, proportional to the cosine of $2\beta$ plus the relative phase of $A_{-1,1}$ with respect to $A_{0,0}$,  from which the relative phase itself can be extracted.

\subsection{Rescaled intensities}

\begin{figure}
    \centering
    \includegraphics[width=\linewidth]{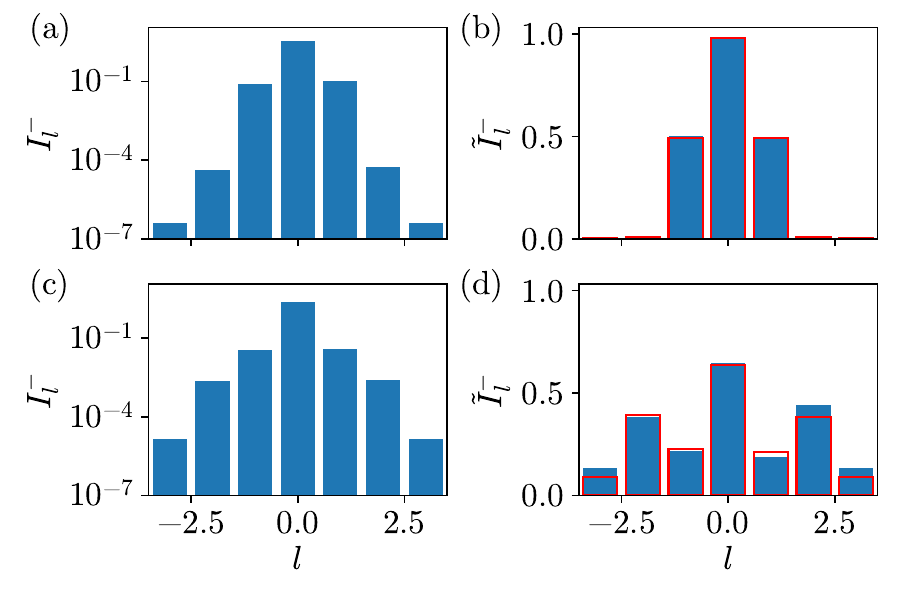}
    \caption{Intensity of $-$ polarized light emitted from the two lowest-lying two-excitation $L=2$ eigenstates in the $D_0=0.6$, $d_0=0.3$ system, as a function of light angular momentum $l$. The upper and lower row corresponds to the ground and first excited state, respectively. The left and right column corresponds to the single-photon intensities without ($I_l^{-}(\ket{\psi})$) and with rescaling ($\tilde{I}_l^{-}(\ket{\psi})$), respectively. The red lines in (b),(d) denote the exact occupation of corresponding orbitals $\ket{\phi^0_L}$.}
    \label{fig:RescaledOutput}
\end{figure}

The information carried by $I_{l}^{\sigma}(\ket{\psi})$ and $I_{l_1l_2}(\ket{\psi})$ is obscured by the suppression of emission into the higher LG modes. This can be seen in Figs.~\ref{fig:RescaledOutput}(a) and (c), which show $I_{l}^{-}(\ket{\psi})$ emitted from the two-excitation ground and excited state, respecitvely. In both cases, the $l=0$ component dominates, and higher-$l$ components are orders of magnitude lower.

But, if we know $f_l^{\sigma}$ (either from single-excitation experiments or simulations) we can look at rescaled output, 
\begin{equation}
    \tilde{I}_{l}^{\sigma}(\ket{\psi})=\frac{I_{l}^{\sigma}(\ket{\psi})}{f_l^{\sigma}}.
    \label{eq:rescaled_oneph}
\end{equation}
It turns out that such a rescaling approximately recovers the orbital populations (at least in the low-lying states composed out of lower-band orbitals), see Fig.~\ref{fig:RescaledOutput}(b). This allows one to differentiate the two-excitation $L=2$ ground state from other two-excitation eigenstates, e.g. the $L=2$ first excited state (Fig.~\ref{fig:RescaledOutput} (d)). Note that the exact occupations $n_{L}^{i}=\braket{\psi |\hat{b}_{iL}^{\dagger} \hat{b}_{iL}|\psi}$, where $\hat{b}_{iL}$ is an annihilation operator for the $i$th orbital at angular momentum $L$, are not the same as in the model state \eqref{eq:laughlin_2part_1d} (by sheer coincidence they resemble rather \eqref{eq:laughlin_2part}), which is expected, as the ground state has imperfect overlap with \eqref{eq:laughlin_2part_1d} (the overlap grows as we lower the lattice constant). The rescaled intensities $\tilde{I}_{l}^{\sigma}$ reflect the orbital populations because the array orbitals in the lower band are primarily composed of the $\ket{-_i}$ atomic orbitals, so emitting a photon is almost equivalent to annihilating a particle from the array orbital (times normalization given by $\sqrt{f_l^{\sigma}}$). A similar rescaling can be performed on the two-photon intensity, 
\begin{equation}
\tilde{I}_{l_1l_2}(\ket{\psi})=\frac{1}{1+\delta_{l_1, l_2}} \frac{I_{l_1l_2}(\ket{\psi})}{f_{l_1}^{-}f_{l_2}^{-}},    
\label{eq:rescaled_twoph}
\end{equation}
which we want to reflect the modulus square of the wavefunction coefficient in front of $\ket{\phi_{L_1}^{0} \phi_{L_2}^{0}}$ ($L_{1,2}=(l_{1,2}+1)~\mathrm{mod}~6$). The factor $\frac{1}{1+\delta_{l_1, l_2}}$ takes into account the $\sqrt{2}$ in the Fock state normalization in case of a doubly occupied orbital, $\ket{\phi_{L}^{0} \phi_{L}^{0}}=\frac{b_{0L}^{\dagger}b_{0L}^{\dagger}}{\sqrt{2}}\ket{\emptyset}$.

\subsection{Signatures of the Laughlin-like state}\label{ssec:signatures}

One can also identify the Laughlin state without the help of $f_l^{\sigma}$. From Eq.~\eqref{eq:laughlin_2part_1d} one can note the following characteristic features of the model state: (i) the equal occupations of the $\ket{\phi^0_0}$ and $\ket{\phi^0_2}$ orbitals, (ii) nonzero occupation of the orbitals $\ket{\phi^0_0}$,$\ket{\phi^0_1}$ and $\ket{\phi^0_2}$ only, (iii) relative phase of $\pi$ between the $\ket{\phi^0_0 \phi^0_2}$ and $\ket{\phi^0_1 \phi^0_1}$ Fock states. In an array eigenstate which has non-perfect overlap with the model state, we expect that the conditions (i)-(iii) hold approximately. We can propose the following characteristic quantities: $R_1(\ket{\psi})=n_{0}^{0}/n_{2}^{0}$, $R_2(\ket{\psi})=n_{3}^{0}/n_{2}^{0}$, $R_2'(\ket{\psi})=n_{5}^{0}/n_{0}^{0}$, $R_3(\ket{\psi})=\arg(C_{02}^{00}/C_{11}^{00})$, with $C_{l_1 l_2}^{ij}$ being a coefficient of the Fock state $\ket{\phi_{l_1}^{i}\phi_{l_2}^{j}}$ in the wavefunction $\ket{\psi}$. For a Laughlin-like state, we would expect $R_1(\ket{\psi})\approx 1$, $R_2\ll 1$, $R_2'\ll 1$, $R_3\approx \pi$. Assuming that $f_1^{-}\approx f_{-1}^{-}$, these characteristics can be translated into the language of output modes, 
\begin{align}
& S_1(\ket{\psi})=\frac{I_{1}(\ket{\psi})}{I_{-1}(\ket{\psi})},\label{eq:S1}\\ 
& S_2(\ket{\psi})=\frac{I_{-2}(\ket{\psi})}{I_{-1}(\ket{\psi})},\label{eq:S2}\\
& S_2'(\ket{\psi})=\frac{I_{2}(\ket{\psi})}{I_{1}(\ket{\psi})},\label{eq:S2prime}\\
& S_3(\ket{\psi})=\arg\left(A_{-1,1}\left(\ket{\psi}\right)/A_{0,0}\left(\ket{\psi}\right)\right).\label{eq:S3}    
\end{align}
For a Laughlin-like state, we would expect $S_1(\ket{\psi})\approx 1$, $S_2(\ket{\psi})$ and $S_2'(\ket{\psi})$ lower than the typical ratio $I_{l_2}/I_{l_1}$ at $|l_2|=|l_1+1|$, and $S_3(\ket{\psi})\approx \pi$.

\begin{figure}
    \centering
    \includegraphics[width=\linewidth]{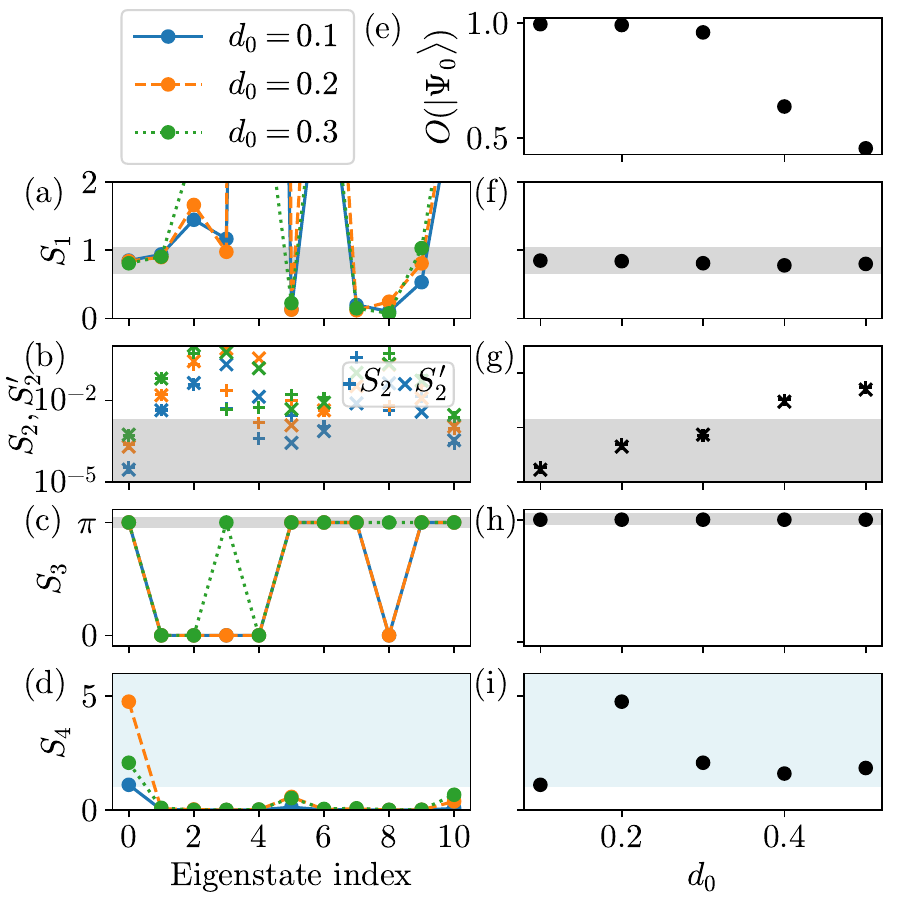}
    \caption{Signatures $S_1-S_4$ (Eqs. \eqref{eq:S1}-\eqref{eq:S4}) of the Laughlin-like states in the output light: (a)-(d) the signatures computed for each two-excitation $L=2$ eigenstate and for three different $d_0$, (e) overlap of the two-excitation $L=2$ ground state with the model Laughlin state \eqref{eq:laughlin_2part_1d} as a function of $d_0$, (f)-(i) signatures computed for the $L=2$ ground state as a function of $d_0$.}
    \label{fig:OutputSignatures}
\end{figure}

Fig. \ref{fig:OutputSignatures} (a)-(c) shows these signatures as a function of eigenstate number (the eigenstates are ordered according to increasing energy) for two-excitation $L=2$ eigenstates at various $d_0$. The gray areas are chosen to contain the two-excitation $L=2$ ground states at $d_0\leq 0.3$ (which have high overlaps with Laughlin states, see Fig.~\ref{fig:OutputSignatures}(e)). In choosing these areas, we also included results for a bigger system, see later text. The numerical values of these areas are $S_1\in [0.65, 1.05]$, $S_2, S_2'\in [0,0.002]$,  $S_3\in [3, 3.3]$. It can be seen that other states of these systems (excited states) sometimes fall within one of the gray areas, but never in all three of them. Therefore, to differentiate the Laughlin state from other two-particle $L=2$ eigenstates, one can compute the signatures and check whether all of them lie within the gray areas.

Additionally, we can introduce a fourth signature based on the $+$ polarized light. We can use the fact that the $|\phi_5^0\rangle$ and $|\phi_0^0\rangle$ orbitals emit $+$ polarized light into the $l=0$ and $l=1$ modes, respectively. Given that for small $d_0$ the former emits more strongly, i.e. $f_0^{+}\gg f_{1}^{+}$ (see Fig. \ref{fig:MinimalSPOutput}), the $+$ polarized light is sensitive to the occupancy of the $\ket{\phi_5^0}$ orbitals. If we measure $I_1^{+}(\ket{\psi})>I_0^{+}(\ket{\psi})$, it implies that $n^{0}_{0}\gg n_{5}^{0}$, i.e. condition on $R_2'$ is fulfilled. Therefore, we define the signature 
\begin{equation}
S_4(\ket{\psi})=\frac{I_1^{+}(\ket{\psi})}{I_0^{+}(\ket{\psi})}.
\label{eq:S4}
\end{equation}
We expect that $S_4>1$ will signify a Laughlin state. This signature is plotted in Fig.~\ref{fig:OutputSignatures}(d), with light blue area denoting $S_4 >1$. It is not perfect, in the sense that at small $d_0$ the Laughlin-like two-excitation ground state falls outside the blue region ($S_4>1$), but no two-excitation excited states fall inside the blue region (i.e. for small $d_0$ it yields false negatives, but not false positives).

It is also instructive to consider higher $d_0$ values, for which the Laughlin-like spectrum becomes distorted, and the overlap of the $L=2$ two-excitation ground state with the Laughlin state falls down (see Fig.~\ref{fig:OutputSignatures}(e)). Fig.~\ref{fig:OutputSignatures}(f)-(i) show the signatures for the $L=2$ two-excitation ground state as a function of $d_0$. Interestingly $S_1$ and $S_3$ remain in the gray areas even when the two-excitation ground state is no longer similar to a Laughlin state, but $S_2$ and $S_2'$ grow with $d_0$ and make it possible to distinguish between Laughlin and non-Laughlin cases. Also, we have $S_4>1$ even for high $d_0$, which means that this time it yields false positives. The reason is that the condition $f_0^{+}\gg f_{1}^{+}$ is no longer satisfied for high $d_0$. Therefore, $S_4$ only is a valid signature of a Laughlin state if $d_0$ is small enough.

\subsection{A bigger system}\label{ssec:bigger}

Similar analysis can be performed for bigger systems, for example $D_0=1.6$. However, here the interpretation of the measurements is less clear, because there is more than one lowest-band orbital at a given array angular momentum $L$, and all orbitals at the same array angular momentum $L$ emit into the same mode. Therefore, it is not possible to distinguish whether the light came, say, from orbital $\ket{\phi_{0}^0}$ or from orbital $\ket{\phi_{0}^1}$. 

\begin{figure}
    \centering
    \includegraphics[width=\linewidth]{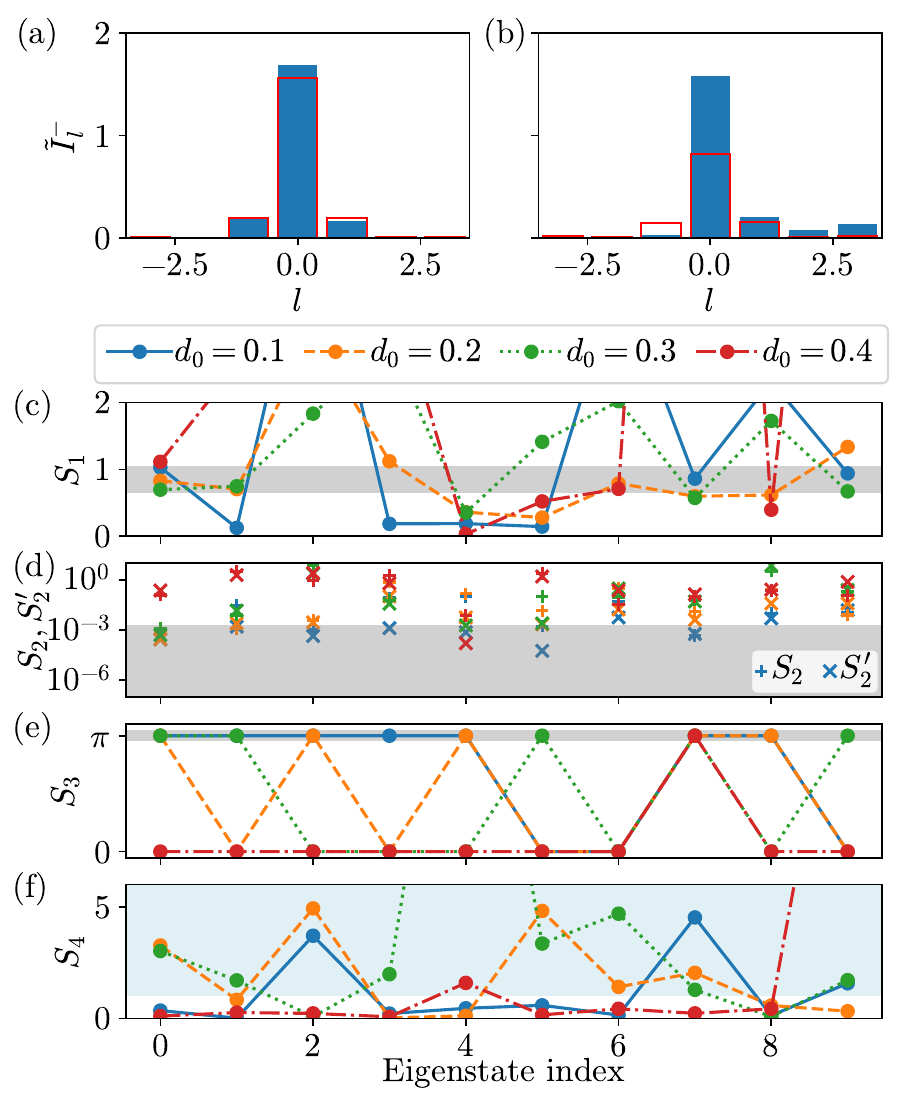}
    \caption{Output light from two-particle $L=2$ eigenstates of $D_0=1.6$ systems. (a), (b) Rescaled single-particle intensities $\tilde{I}^-_{l}(\ket{\psi})$ of the light emitted from the ground state and first excited state, respectively, as a function of angular momentum $l$ of the light, for a system with $d_0=0.1$ and $V_\mathrm{harm0}=3$. The red lines denote the exact occupation of corresponding orbitals $\ket{\phi^0_L}$. (c)-(f) Signatures $S_1-S_4$ (Eqs.~\eqref{eq:S1}-\eqref{eq:S4}) of the Laughlin-like states in the output light from lowest 10 eigenstates. The confining potentials used are $V_\mathrm{harm0}=3$ for $d_0=0.1$ and $V_\mathrm{harm0}=10$ otherwise.}
    \label{fig:OutputSignaturesBigger}
\end{figure}

This is visible for example if we look at Figs.~\ref{fig:OutputSignaturesBigger}(a), (b), showing the rescaled single-photon intensities \eqref{eq:rescaled_oneph} of light emitted from a two-excitation $L=2$ ground state and first excited state, respectively, in a $d_0=0.1$, $D_0=1.6$ system with $V_\mathrm{harm0}=3$. These results are compared with the exact occupations of the lowest orbitals at given array angular momentum $L=l+1\mod 6$ (i.e. $\ket{\phi_{L}^0}$). While for the two-excitation ground state the rescaled intensities seem to be a good proxy for occupations, for the first excited two-excitation state they are not. The reason is the contribution of higher orbitals. For example, the first excited state has a strong contribution of the $\ket{\phi_0^0\phi_0^1}$ Fock state, which we cannot differentiate from $\ket{\phi_0^0\phi_0^0}$.

Nevertheless, even taking into account that the rescaled intensities do not have to be a good approximation of occupations, we can still see that the rescaled angular-momentum-resolved emission spectra of the two-excitation ground and excited state differ qualitatively, for example with the latter being visibly asymmetric. That is, we have $S_1\ll 1$. This suggests that we can still use signatures to differentiate the Laughlin-like states from other states. 

The signatures for $D_0=1.6$ are plotted in Fig.~\ref{fig:OutputSignaturesBigger} (c)-(f), with the gray and blue regions the same as in Fig.~\ref{fig:OutputSignatures}. Here, again, the systems $d_0=0.1, 0.2, 0.3$ display a clear Laughlin-like two-excitation $L=2$ ground state while $d_0=0.4$ does not. The behavior of the signatures is similar as in Fig.~\ref{fig:OutputSignatures}, with three key differences: (i) $S_3$ clearly differentiates the non-Laughlin case of $d_0=0.4$ from the Laughlin-like cases, (ii) $S_4$ becomes useless (or at least much less useful) because now false positives are pretty frequent, and (iii) there are two-excitation states other than ground states (such as state 7 at $d_0=0.1$) for which all four signatures $S_1$, $S_2$, $S_2'$, $S_3$ lie within the gray regions, which means that these signatures combined can now yield false positives. The presence of such states should be ruled out based on other considerations (e.g. energy).

To sum up, we have shown that in the relevant regime, optical signatures can distinguish the array analog of the two-particle Laughlin ground state from all other two-excitation eigenstates in the nanoring case $D_0=0.6$ and from most of the other eigenstates in bigger systems. The question is now: how can such a two-excitation Laughlin-like ground state be prepared? And can one tell from optical signatures that the actual state generated is close to a Laughlin-like state?

\section{Driving}\label{sec:driving}
\subsection{General remarks about driven system}\label{ssec:driving_general}

\begin{figure}
    \centering
    \includegraphics[width=\linewidth]{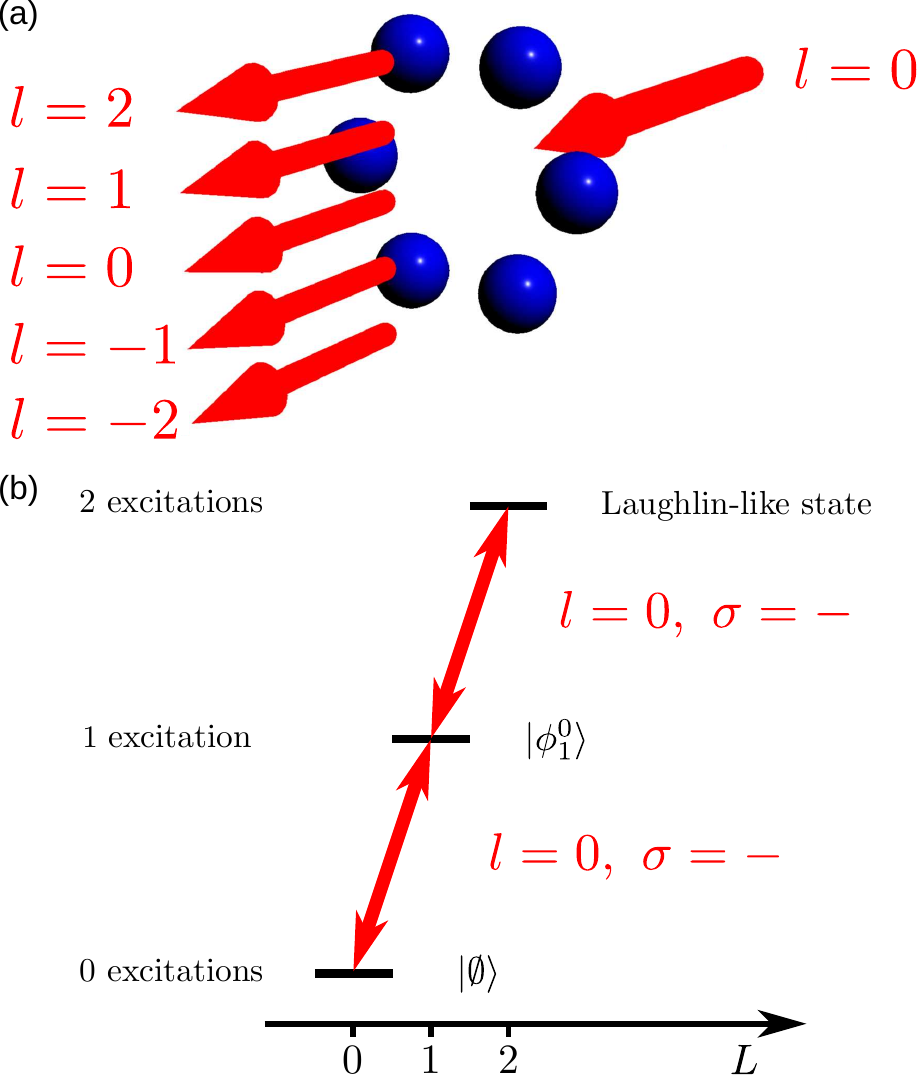}
    \caption{The driving scheme. (a) A schematic picture of the driven system. It is illuminated by $l=0$ light, but emits into modes with various angular momenta. (b) The energy levels involved in the driving process.}
    \label{fig:DrivingScheme}
\end{figure}

We established in Sec.~\ref{sec:laughlinlike} that the two-excitation Laughlin-like state responds strongly to light. We can then suspect that it can be efficiently excited by driving. Here we investigate such a possibility. We consider driving with a $-$ polarized light with angular momentum $l=0$ (Fig.~\ref{fig:DrivingScheme}(a)). The ideal driving scheme, similar to the one used in \cite{umunculiar2014probing,clark2020observation}, is as follows: the first absorbed photon will move the system from the no-excitation state $\ket{\emptyset}$ to a single-excitation $L=1$ state $\ket{\phi_1^0}$, while the second photon will move it to the Laughlin-like state due to the contribution of the $\ket{\phi_1^0\phi_1^0}$ Fock state to the Laughlin-like state (Fig.~\ref{fig:DrivingScheme}(b)). This scheme assumes that the resonance between the $\ket{\emptyset}$ and $\ket{\phi_1^0}$ states occurs at similar detuning as the resonance between $\ket{\phi_1^0}$ and the two-excitation Laughlin-like state. This expectation is reasonable, as in the LLL the interaction energy of the Laughlin state vanishes, and we expect that the same should be approximately true in arrays, i.e. that the energy of the Laughlin-like state is approximately twice the energy of the orbital $\ket{\phi_1^0}$.

To see to what extent this picture is true, we consider specifically the Hamiltonian in the presence of driving, which in the rotating wave approximation is given by

\begin{multline}
    \hat{\tilde{H}}_\mathrm{tot}=\hat{H}-\Delta\sum_{i\sigma} \hat{a}^{\dagger}_{i\sigma}\hat{a}_{i\sigma} \\- \sum_{i} \left(\Omega^{*}(\mathbf{r}_i)
 \hat{a}_{i-}+
    \Omega (\mathbf{r}_i) \hat{a}_{i-}^{\dagger}\right),
\end{multline}
where $\Delta=\omega-\nu$, with $\nu$ being the driving frequency, and $\Omega(\mathbf{r})$ is a Rabi frequency. In general, $\Omega$ can depend on position, but we assume that it is uniform, i.e. the system is driven by a plane wave (given the small system size compared to $\lambda_A$, this can also approximate the case of the $l=0$ Laguerre-Gauss mode).

In the weak driving approximation, the system, although open, can still be described with a wavefunction. We write it as
\begin{equation*}
\ket{\psi_\mathrm{steady}}=c_0\ket{\psi_\mathrm{steady}^{(0)}}+c_1\ket{\psi_\mathrm{steady}^{(1)}}+c_2\ket{\psi_\mathrm{steady}^{(2)}}+\dots
\label{eq:steady_npart_decomposition}
\end{equation*}
where $\ket{\psi_\mathrm{steady}^{(n)}}$ is a normalized wavefunction in an $n$-excitation space, and $c_n$ is an overall coefficent of this contribution. We assume that  $|c_0|\gg |c_1| \gg |c_2|$. Then, we determine $c_n$ and  $\ket{\psi_\mathrm{steady}^{(n)}}$ from  $c_{n-1}$ and  $\ket{\psi_\mathrm{steady}^{(n-1)}}$ by demanding that the wavefunction up to $n$ excitations is a steady state,
\begin{multline}
    i\frac{\mathrm{d}}{\mathrm{d}t} c_n\ket{\psi_\mathrm{steady}^{(n)}}=-\Delta\sum_{i\sigma} \hat{a}^{\dagger}_{i\sigma}\hat{a}_{i\sigma} c_n\ket{\psi_\mathrm{steady}^{(n)}}+ \\
    \hat{H} c_n\ket{\psi_\mathrm{steady}^{(n)}}
    -\Omega\sum_{i} c_{n-1}\hat{a}_{i-}^{\dagger} \ket{\psi_\mathrm{steady}^{(n-1)}}\stackrel{!}{=}0,
    \label{eq:weakdriving}
\end{multline}
and keeping $c_{n-1}$ and  $\ket{\psi_\mathrm{steady}^{(n-1)}}$ constant because $|c_{n-1}|\gg |c_n|$. 

\begin{figure}
    \centering
    \includegraphics[width=\linewidth]{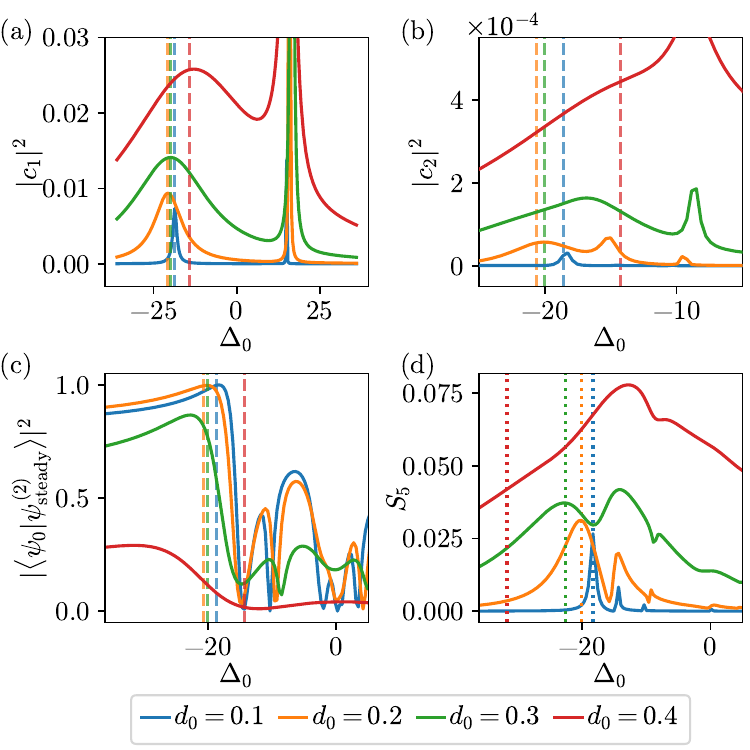}
    \caption{Driving the $D_0=0.6$ system with $\Omega=0.1 \Gamma_0$ Rabi frequency. (a) Probability of having exactly one atomic excitation in the system, as a function of detuning. (b) Probability of having exactly two atomic excitations in the system, as a function of detuning. (c) Squared overlap of the normalized two-excitation component of the steady state with the two excitation $L=2$ ground state of the Hermitian part of Hamiltonian \eqref{eq:ham}. (d) The signature $S_5$ \eqref{eq:S5} of the maximum overlap. The dashed vertical lines in (a)-(c) denote the energies of the lower-band $L=1$ orbital, while the dotted vertical lines in (d) denote the maxima of overlap from (c).
 }
    \label{fig:DrivingMain}
\end{figure}

\begin{figure}
    \centering
    \includegraphics[width=\linewidth]{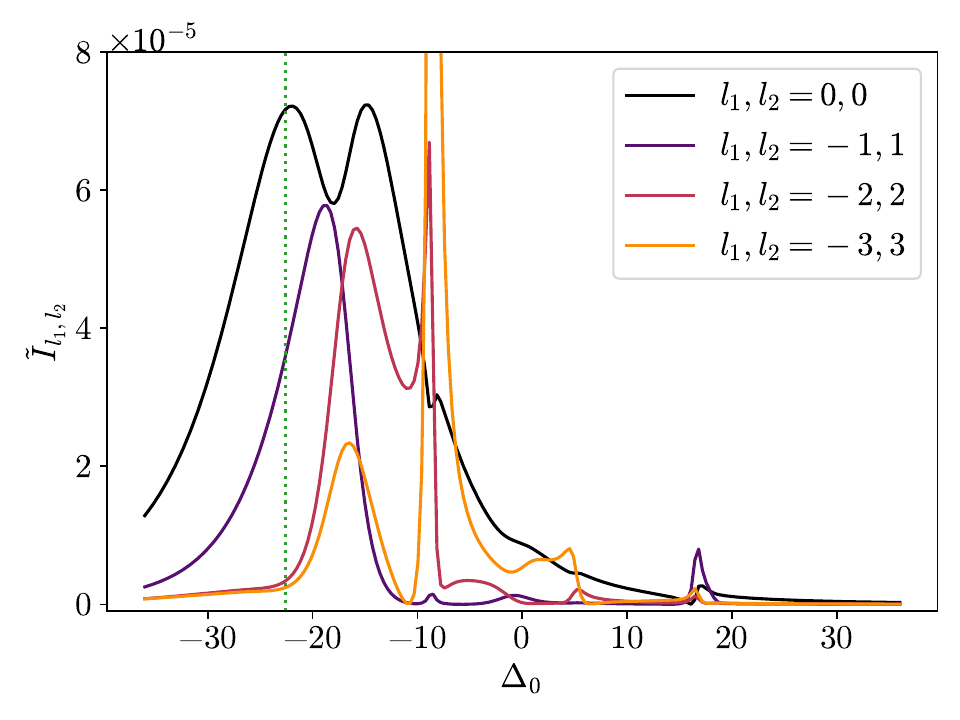}
    \caption{Rescaled two-photon intensities emitted by the steady state of the $D_0=0.6$, $d_0=0.3$ system, as a function of detuning. The green dotted vertical line marks the position of maximum overlap of the normalized two-excitation component with the two excitation $L=2$ ground state of the Hermitian part of Hamiltonian \eqref{eq:ham}.}
    \label{fig:RenormalizedDriving}
\end{figure}

\subsection{Steady state at $D_0=0.6$}

The results for $D_0=0.6$ are shown in Fig.~\ref{fig:DrivingMain}. In particular, we plot the fraction $|c_n|^2$ of the steady state in the $n$-excitation subspace as a function of detuning (Fig.~\ref{fig:DrivingMain}(a),(b)). To compare the results at different $d_0$, we have introduced the rescaled detuning $\Delta_0=\frac{d^3k_A^3}{\Gamma_0} \Delta$ on the $x$-axis. The resonance between the atomic ground state and the lower-band single-excitation $L=1$ orbital for $d_0\leq 0.3$ occurs at similar $\Delta_0$, around $\Delta_0=-20$ regardless of the lattice constant (at $d_0=0.4$ it moves to around $\Delta_0=-13$). In Fig.~\ref{fig:DrivingMain}(a), the energies of the lower-band $L=1$ orbital are denoted by dashed vertical lines. The first peak of the $|c_1|^2$ curves is located close to the detuning equal to the energy of the orbital (the discrepancy is apparent only at $d_0=0.4$ and is there because the energies are obtained from diagonalization of the Hermitian part of the Hamiltonian, not the full non-Hermitian Hamiltonian). Similarly, the second peak of $|c_1|^2$ corresponds to the resonance with the upper-band $L=1$ orbital. The fact that the second peak is narrower reflect the smaller decay rate of the upper-band orbital.

As mentioned in Subsection \ref{ssec:driving_general}, we expect that the energy of the Laughlin-like state is approximately twice the energy of orbital $\ket{\phi_1^0}$, thus a resonance between the $\ket{\phi_1^0}$ state and the Laughlin-like state should occur at similar detuning as the resonance between the $\ket{\emptyset}$ and $\ket{\phi_1^0}$ states. Fig.~\ref{fig:DrivingMain}(b), showing the fraction $|c_2|^2$ in the two-excitation subspace, suggests that this is indeed the case for small $d_0$. At $d_0=0.1$ and $d_0=0.2$ (blue and orange curves, respectively), the first peak of $|c_2|^2$ occurs slightly to the right of the dashed vertical lines of respective color, which again denote the orbital energies. At $d_0=0.3$, this peak is no longer discernible from the second peak.

At the first peak of $|c_2|^2$ at small $d_0$, the normalized two-excitation contribution to the steady state indeed resembles the two-excitation Laughlin-like ground state of the Hermitian part of the static Hamiltonian. This is confirmed in Fig.~\ref{fig:DrivingMain}(c), showing that the overlap between these states achieves values close to 1 near the dashed vertical lines (again denoting the energies of the orbitals). Interestingly, the overlap is still quite high in the $d_0=0.3$ case even though the $|c_2|^2$ peak in Fig.~\ref{fig:DrivingMain}(b) is not discernible. In contrast, at $d_0=0.4$ the maximum overlap becomes low. This, combined with the fact that at $d_0=0.4$ the static two-excitation $L=2$ ground state has quite low overlap with the model Laughlin state, indicates that the range of $d_0$ suitable for generating Laughlin-like states is $d_0\sim 0.3$ or
lower.

One can pinpoint the resonance between the $\ket{\phi_1^0}$ state and the two-excitation Laughlin-like ground state more precisely by investigating the ratio $|c_2|^2/|c_1|^2$. This quantity can be associated with a measurable signature. Let us define analogs of the observables \eqref{eq:I1p}-\eqref{eq:A2p} from Section \ref{sec:output}, 
\begin{align}
&I_l^{\sigma}( \ket{\psi_\mathrm{steady}} )= (\delta_{l,0}\delta_{\sigma,-}+\delta_{l,2}\delta_{\sigma,+})|c_1|^2 I_l^{\sigma}( \ket{\psi_\mathrm{steady}^{(1)}} )\nonumber\\
& +|c_2|^2 I_l^{\sigma}( \ket{\psi_\mathrm{steady}^{(2)}} ),\label{eq:I1pdriven} \\
&I_{l_1, l_2}( \ket{\psi_\mathrm{steady}} )= |c_2|^2 I_{l_1, l_2}( \ket{\psi_\mathrm{steady}^{(2)}} )\label{eq:I2pdriven}, \\
&A_{l_1, l_2}( \ket{\psi_\mathrm{steady}} )= c_2 A_{l_1, l_2}( \ket{\psi_\mathrm{steady}^{(2)}} ) \label{eq:A2pdriven}.    
\end{align}
The delta factors in \eqref{eq:I1pdriven} are there because the intensity is a sum of contribution from single- and two-excitation subspaces, with the the former containing only the $l=0$ $\sigma=-$ and $l=2$ $\sigma=+$ modes allowed by the angular momentum conservation. Using Eqs.~\eqref{eq:I1pdriven}, \eqref{eq:I2pdriven}, we can define a quantity depending on $|c_2|^2/|c_1|^2$,
\begin{multline}
S_5(\ket{\psi_\mathrm{steady}})=\frac{I_{0,0}(\ket{\psi_\mathrm{steady}})}{I_0^-(\ket{\psi_\mathrm{steady}})} \\ \approx  \frac{|c_2|^2 I_{0,0}(\ket{\psi_\mathrm{steady}^{(2)}})}{|c_1|^2 I_0^-(\ket{\psi_\mathrm{steady}^{(1)}})}    .
\label{eq:S5}
\end{multline}
The signature \eqref{eq:S5} is plotted in Fig.~\ref{fig:DrivingMain}(d). The detunings corresponding to maximum overlap are denoted by the vertical dotted lines. It can be seen that for $d_0$ up to 0.3 the maxima of $S_5$ correspond to maxima of overlap. At $d_0=0.4$, when the maximum overlap is small, this correspondence breaks down.

\subsection{Detecting the Laughlin state with rescaled intensities}

Signature $S_5$ tells the experimenter at which value of detuning to look for the Laughlin state. But can one deduce from observables \eqref{eq:I1pdriven}-\eqref{eq:A2pdriven} that at this value the two-excitation component of the steady state is indeed Laughlin-like?

From \eqref{eq:I1pdriven} one can see that one should avoid using $I_0^{-}( \ket{\psi_\mathrm{steady}} )$, as the light from the Laughlin-like state will be obscured by the emission from the single-excitation subspace. Instead, one option is to use two-photon intensities, which only contain contributions from the two-excitation subspace. We rescale them as in Eq.~\eqref{eq:rescaled_twoph}.  The results for $d_0=0.3$ are shown in Fig.~\ref{fig:RenormalizedDriving}. The dotted green line is the detuning corresponding to maximum overlap. 

Because the static two-excitation $L=2$ ground state is close to $(\ket{\phi_{1}^{0}\phi_{1}^{0}}-\ket{\phi_{0}^{0}\phi_{2}^{0}})/\sqrt{2}$ (as indicated in Fig.~\ref{fig:RescaledOutput}(b)) one might expect that at the intersection with the dotted green line, the two rescaled two-photon intensities $\tilde{I}_{0,0}(\ket{\psi_\mathrm{steady}})$ and $\tilde{I}_{-1,1}(\ket{\psi_\mathrm{steady}})$ will have equal magnitudes. This is not the case: the former is about twice as strong as the latter, as indicated by the values of the black and purple curves where they intersect the dotted green line. This accounts for the overlap with static two-excitation $L=2$ ground state not being perfect. In any case, we still observe that the rescaled intensities $\tilde{I}_{0,0}(\ket{\psi_\mathrm{steady}})$ and $\tilde{I}_{-1,1}(\ket{\psi_\mathrm{steady}})$ are much stronger than $\tilde{I}_{-2,2}(\ket{\psi_\mathrm{steady}})$ and $\tilde{I}_{-3,3}(\ket{\psi_\mathrm{steady}})$, suggesting that the $\ket{\phi_{5}^{0}\phi_{3}^{0}}$, $\ket{\phi_{4}^{0}\phi_{4}^{0}}$ Fock states have at most a small contribution to the steady state, as expected for a Laughlin-like state. We note that the $\tilde{I}_{-2,2}(\ket{\psi_\mathrm{steady}})$ and $\tilde{I}_{-3,3}(\ket{\psi_\mathrm{steady}})$ intensities have minima near the green dotted vertical line, i.e. maximum of overlap. This is, however, just a coincidence. While the small values of $\tilde{I}_{-2,2}(\ket{\psi_\mathrm{steady}})$ and $\tilde{I}_{-3,3}(\ket{\psi_\mathrm{steady}})$ intensities are a hallmark of a Laughlin-like state, at other values of $d_0$ the exact location of the minimum does not coincide with maximum overlap.

\subsection{Detecting the Laughlin state using signatures}

\begin{figure}
    \centering
    \includegraphics[width=\linewidth]{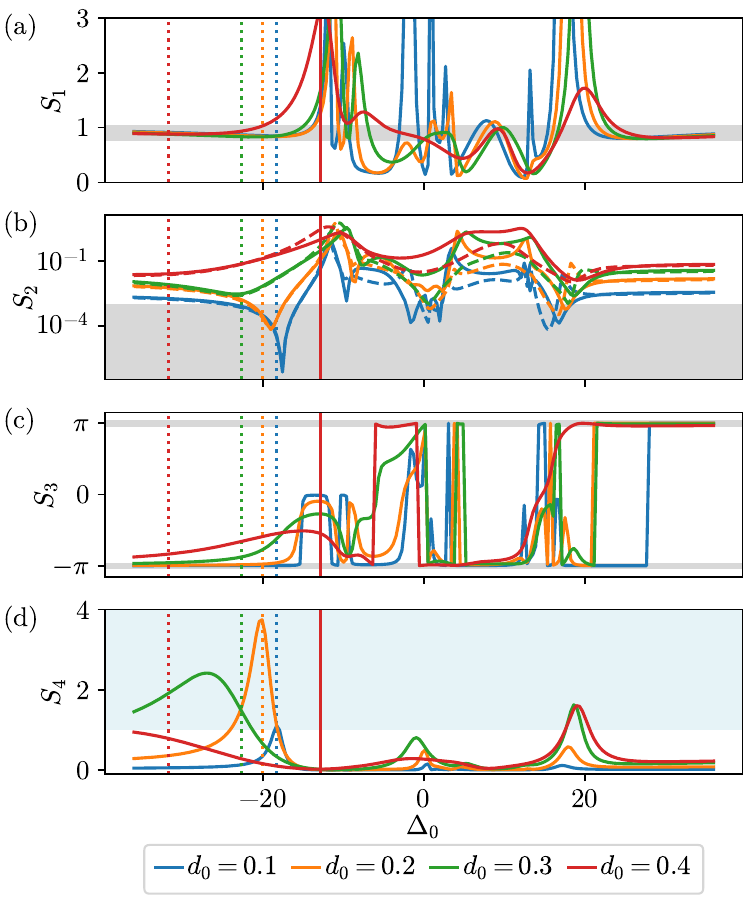}
    \caption{Signatures \eqref{eq:S1}-\eqref{eq:S4} of the Laughlin-like state as a function of detuning in $D_0=0.6$ systems with various $d_0$. Solid and dashed lines in (c) denote $S_2$ and $S_2'$, respectively. Dotted vertical lines denote the location of maximum overlap between $\ket{\psi_\mathrm{steady}^{(2)}}$ and the static $L=2$ ground state. The red solid vertical line denotes the first maximum of $S_5$ for $d_0=0.4$. In (c), the solid and dashed lines represent $S_2$ and $S_2'$, respectively.}
    \label{fig:DrivingSignatures}
\end{figure}

Can one identify the Laughlin state using signatures $S_1(\ket{\psi_\mathrm{steady}})$-$S_4(\ket{\psi_\mathrm{steady}})$ given by Eqs. \eqref{eq:S1}-\eqref{eq:S4}? Fig.~\ref{fig:DrivingSignatures} shows these quantities as a function of detuning. It can be seen that for $d_0=0.1$, $0.2$, $0.3$ the signatures at maximum overlap (detectable as maximum of $S_5(\ket{\psi_\mathrm{steady}})$), represented by dotted vertical lines, either lie in the gray/blue areas or are close to them, confirming the presence of the two-excitation Laughlin-like state.

Applying the same procedure at $d_0=0.4$ yields a negative result. As one can remember from Fig.~\ref{fig:DrivingMain}(d), the first maximum of $S_5$ does not lie at the maximum overlap, but rather near $\Delta\approx - 13$ (this detuning is indicated in Fig.~\ref{fig:DrivingSignatures} as solid red vertical line). Therefore, the experimenter would look for the Laughlin state there, rather than in the true maximum of overlap. From Fig.~\ref{fig:DrivingSignatures} one can see that at the intersection with the solid red vertical line, the signatures are far from the gray/blue boxes, showing that the two-excitation component of the steady state is not similar to the Laughlin state. The actual maximum of overlap (red dotted line) is not detectable using our method.

\subsection{A bigger system}

\begin{figure}
    \centering
    \includegraphics[width=\linewidth]{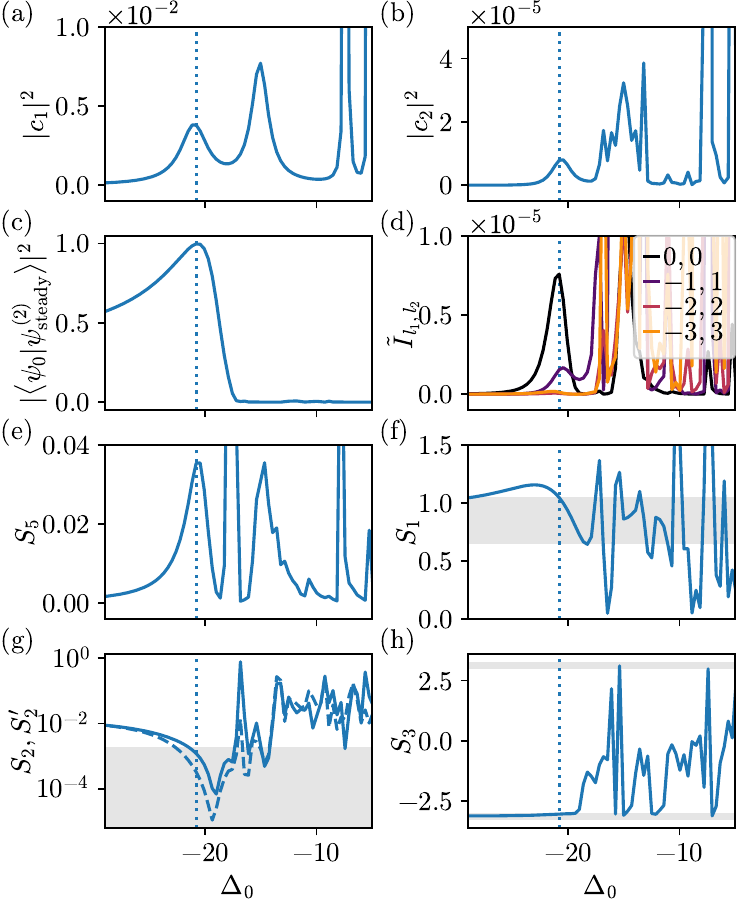}
    \caption{Driving the $D_0=1.6$, $d_0=0.1$, $V_\mathrm{harm0}=3$ system with spatially uniform Rabi frequency $\Omega=0.1\Gamma_0$. (a), (b) the square modulus of the 1- and 2-excitation coefficients $c_1$, $c_2$, respectively. (c) Squared overlap between the two-excitation part of steady state and the static two-excitation $L=2$ ground state. (d) Rescaled two-photon intensity $\tilde{I}_{l_1,l_2}$. (e)-(h) Signatures $S_5$, $S_1$, $S_2$ (with $S_2'$) and $S_3$, respectively  (Eqs.~\eqref{eq:S1}-\eqref{eq:S3} and \eqref{eq:S5}).  Solid and dashed lines in (g) denote $S_2$ and $S_2'$, respectively. The dotted vertical lines in all the subplots denote the position of the maximum of overlap.}
    \label{fig:DrivingSignaturesBigger}
\end{figure}

 Let us now consider an example of a bigger system: a $D_0=1.6$, $d_0=0.1$ system with $V_\mathrm{harm0}=3$. The results are shown in Fig.~\ref{fig:DrivingSignaturesBigger}. We can again  pinpoint the vicinity of maximum overlap  (Fig.~\ref{fig:DrivingSignaturesBigger}(c)) as the maximum of signature \eqref{eq:S5} (Fig.~\ref{fig:DrivingSignaturesBigger}(e)), and then observe that at this point the rescaled two-photon emission is dominated by $\tilde{I}_{0,0}$ and $\tilde{I}_{-1,1}$  (Fig.~\ref{fig:DrivingSignaturesBigger}(d)), and that all the signatures \eqref{eq:S1}-\eqref{eq:S3} lie within the gray regions  (Fig.~\ref{fig:DrivingSignaturesBigger}(f)-(h); we recall from Sec.~\ref{sec:output} that signature \eqref{eq:S4} is not useful for systems bigger than 6 atoms). As noted before, the interpretation of these signatures is more ambiguous, as multiple lower-band orbitals emit into the same mode. Still, the signatures offer testable predictions consistent with a scenario that a Laughlin state is created. More results on the $D_0=1.6$ systems can be found in Appendix~\ref{app:bigger}.

 Finally, let us comment on applicability of our scheme to the case of more than two particles. In principle, it is possible to use the weak driving setup for more particles. We have verified that one can drive the 3-particle Laughlin-like ground state in a $D_0=0.6$ system with $l=1$ light. However, one should note that the population of each $N_\mathrm{part}$ sector reduces strongly with increasing $N_\mathrm{part}$, making it harder to measure the relevant quantities. Another issue is the identification of the Laughlin state: to unambiguously detect the $3$-particle FQH state using methods presented in this work, we would have to study several three-photon amplitudes and their relative phases, which would be even more challenging experimentally than the two-particle case. Therefore, our method serves as a proof-of-concept of realizing Laughlin states in atom arrays, but is not scalable to larger particle numbers.

In summary, we have shown that the two-particle Laughlin-like states are achievable by driving, and that the effects of its presence show up in the rescaled two-photon intensities, as well as the signatures defined in Sec. \ref{sec:output}.

\section{Experimental considerations}\label{sec:experiment}

While creating deep subwavelength atom arrays remains a challenge, significant experimental effort is directed towards this goal \cite{wang2018dark, rui2020subradiant,srakaev2023subwavelength, du2024atomic}. One kind of proposals is to use a shorter-wavelength transition for trapping the atoms and a longer-wavelength one to perform optical experiments. For example, if bosonic strontium atoms are excited to a metastable $^3P_0$ state, they can be trapped with a magic wavelength $\lambda_\mathrm{trap}=412.8~\mathrm{nm}$, and quantum optics can be studied with $^3P_0$-$^3D_1$ transition with $\lambda_A=2.6~\mu \mathrm{m}$ transition \cite{olmos2013long} with decay rate $\Gamma_0=290 \times 10^3 s^{-1}$ \cite{zhou2010magic}. Considering the case of trapping near diffraction limit, with tweezers spaced $\lambda_\mathrm{trap}$ apart, and noting that that $\lambda_\mathrm{trap}/\lambda_A\sim 0.16$, we get  $d_0\sim 0.27$, suitable for realization of two-particle Laughlin-like states in the nanoring system or bigger systems with strong harmonic confinement potential, as in Figs. \ref{fig:Overlap}(b) and (c). 

One could also envision other platforms, such as excitons in Moir\'e materials \cite{huang2024collective}, where the Moir\'e pattern naturally gives rise to strongly sub-wavelength lattice constants. We also note that our results may be useful for a Rydberg atom platform, where the Hamiltonian is a $d_0\rightarrow 0$ limit of our model \cite{weber2022experimentally}. The results for the minimal nanoring systems indicate that only six Rydberg atoms would be sufficient to demonstrate some properties of the Laughlin states in experiment. A similar system of three atoms was already created \cite{lienhard2020realization}.

\section{Summary and conclusions}
In this work, we have demonstrated that atom arrays with topological bands can exhibit Landau-level-like single-particle wavefunctions, which when combined with the native onsite hard-core interaction, gives rise Laughlin-like states of few atomic excitations. The problem of of the divergence of the dispersion relation near the light cone is circumvented by choosing small lattice constants and small system sizes. The information about the Laughlin-like nature of the two-excitation ground states, in particular the orbital composition of the state and the relative phases of Fock states, is encoded in the Laguerre-Gauss mode decomposition of the output light. The Laughlin-like state can be driven with zero-angular-momentum light, and the output-light signatures proposed by us allow for detecting its presence (less ambiguously in the 6-site systems, more ambiguously in bigger systems).

We regard this work primarily as a conceptual inquiry into few-body effects in an atomic array, a first step in a larger program to realize and understand many-body, strongly correlated phenomena in such systems. Indeed, there are a number of natural questions that emerge from this work. For example, what is the role of long-range interactions and divergent bands, and is it possible to realize a thermodynamic phase featuring topological order? Are there ways to image and control the system with single-site resolution, necessary to measure correlation functions and pin anyonic excitations? Is it possible to observe the whole branch of edge excitations and confirm that they obey fractional exclusion statistics? And can we use the fundamental features of the arrays, such as dissipation and long-range interaction, as useful tools to engineer and probe states?

Separately, it would be interesting to explore with experiments the physics as described in our work, even at the few-body level. While the experiments would certainly be challenging, estimations presented in Sec. \ref{sec:experiment} suggest that our system could in principle be realized with atoms in optical tweezer setup.

\begin{acknowledgments}
The authors acknowledge funding from European Union, MSCA Fellowship no. 101145886 (QUINTO). DEC acknowledges support from the European Union, under European Research Council grant agreement No 101002107 (NEWSPIN) and EIC Pathfinder Grant No 101115420 (PANDA); the Government of Spain (Severo Ochoa Grant CEX2019-000910-S [MCIN/AEI/10.13039/501100011033]); QuantERA II project QuSiED, co-funded by the European Union Horizon 2020 research and innovation programme (No 101017733) and the Government of Spain (European Union NextGenerationEU/PRTR PCI2022-132945 funded by MCIN/AEI/10.13039/501100011033); Generalitat de Catalunya (CERCA program and AGAUR Project No. 2021 SGR 01442); Fundació Cellex, and Fundació Mir-Puig.
\end{acknowledgments}

\section*{Data availability statement} 
The code and data supporting this publication is available at \cite{data}.

\appendix

\section{Restoring rotational symmetry of the flakes}\label{app:rotsymm}

With hopping given by the Green's function \eqref{eq:GreensFunction} the flakes, although hexagonal, do not possess six-fold rotational symmetry. Rather, they are invariant with respect of rotation combined with phase shift, akin to magnetic rotation. 

The reason lies in the angle-dependent factors in $G_{+-}$, $G_{-+}$ in \eqref{eq:GreensFunction}. Consider a pair of sites $i,j$, connected with the vector $\mathbf{r}$ whose angular coordinate is $\phi$. A rotation $R$ by $\pi/3$, transforms them into sites $R(i)$, $R(j)$. These sites are connected by a vector $R(\mathbf{r})$ with an angular coordinate $\phi+\pi/3$, which leads to the fact that $G_{+-}(R(\mathbf{r}))$ has an additional phase of $e^{2i\pi/3}$ compared to $G_{+-}(\mathbf{r})$. 

To get rid of this phase, we define new creation operators $\hat{\tilde{a}}^{\dagger}_{i \sigma}$. For sites $i$ in  ``sector 0'' of Fig.~\ref{fig:SystemPicture}(a), we have $\hat{\tilde{a}}^{\dagger}_{i \sigma}=\hat{a}^{\dagger}_{i \sigma}$. For a site $R^{n}(i)$ that can be obtained by applying the rotation $R$ $n$ times to a site $i$ from sector 0, we define $\hat{\tilde{a}}^{\dagger}_{R^n(i) \pm}=e^{\pm in\pi /3}\hat{a}^{\dagger}_{R^n(i) \pm}$. The Hamiltonian \eqref{eq:hband} written in terms of $\hat{\tilde{a}}^{\dagger}_{i \sigma}$ becomes invariant to $\pi/3$ rotations. Note that the transformation acts trivially on ``sector 0'', so the phase convention mentioned in Sec.~\ref{sec:modelstates} (zero phase in ``sector 0'') applies for orbitals written in terms of both $\hat{\tilde{a}}^{\dagger}_{i \sigma}$ and  $\hat{a}^{\dagger}_{i \sigma}$ operators.

\section{Stability of LLL-like strucutre and Laughlin-like states}\label{app:phasediagrams}
\begin{figure}
    \centering
    \includegraphics[width=\linewidth]{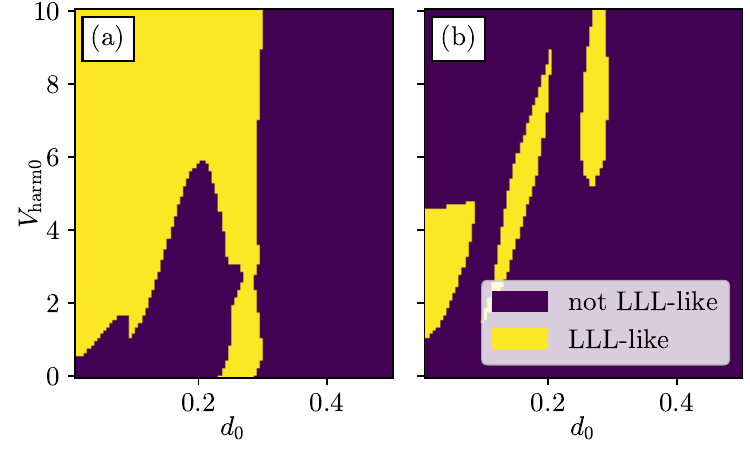}
    \caption{The range of validity of the LLL-like structure of the eigenstates, as defined in the main text, for (a) $m_\mathrm{max}=4$ and (b) $m_\mathrm{max}=9$. The system size is $D_0=2.1$.}
    \label{fig:PhaseDiagramsLLLLike}
\end{figure}

In this Appendix, we study how stable are the single-particle  LLL-like states and few-excitation  Laughlin states with respect to changes in the lattice constant and confining potential. We define the single-excitation eigenstates to have a ``LLL-like structure'' if their radius grows with angular momentum. Let us change the notation and label the single-excitation array eigenfunctions (orbitals) with the continuum angular momentum $m$ of the corresponding continuum LLL orbital $\phi_L^i\equiv \phi_{m=6i+L}$. Because of the finite size of our systems, the growth of radius with $m$ will stop at some point. Therefore, we set some cutoff $m_\mathrm{max}$ and check whether the radius grows with $m$ up to $m_\mathrm{max}$. That is, we say that the eigenstate structure is ``LLL-like'' up to $m_\mathrm{max}$ if for every $m_1<m_2\leq m_\mathrm{max}$ we have $\braket{\phi_{m_1}|\sum_i r_i \hat{n}_i|\phi_{m_1}} <\braket{\phi_{m_2}|\sum_i r_i \hat{n}_i|\phi_{m_2}}$. The parameter range where such condition is fulfilled obviously decreases with increasing $m_\mathrm{max}$, therefore one should choose this parameter based on what one wants to achieve. If one just wants to focus on the Laughlin-like $N_\mathrm{part}$-excitation ground state, $m_\mathrm{max}=2(N_\mathrm{part}-1)$ is sufficient, but the more edge excitations we want to include, the higher $m_\mathrm{max}$ we need, and the smaller is the parameter range where the LLL-like structure exists. Fig.~\ref{fig:PhaseDiagramsLLLLike} shows this range (in the space of $d_0$ and $V_\mathrm{harm0}$) for a $D_0=2.1$ system for $m_\mathrm{max}=4$, sufficient for the $N_\mathrm{part}=3$ Laughlin-like ground state (Fig.~\ref{fig:PhaseDiagramsLLLLike}(a)) and $m_\mathrm{max}=9$, sufficient for all $N_\mathrm{part}=3$ edge excitations with $\Delta L'\leq 6$ (Fig.~\ref{fig:PhaseDiagramsLLLLike}(b)). It can be seen that the former is pretty broad, while the later only consists of isolated islands. From Fig.~\ref{fig:PhaseDiagramsLLLLike}(a), one can also notice that increasing the confining potential increases the maximum available lattice constant for which the structure is still LLL-like.

\begin{figure}
    \centering
    \includegraphics[width=\linewidth]{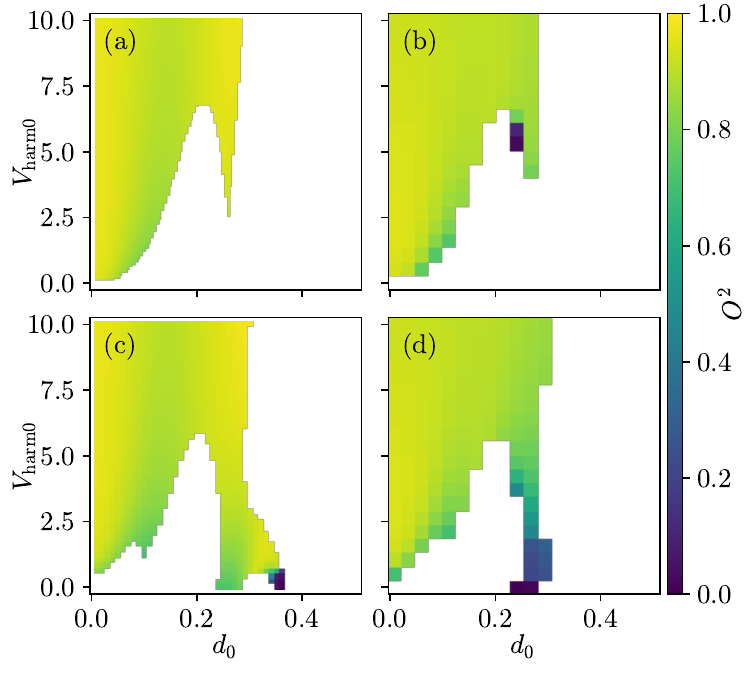}
    \caption{Overlap between model Laughlin ground state and the ground state of the corresponding array angular momentum sector of the Hermitian part of the array Hamiltonian \eqref{eq:ham}. Lower and upper rows correspond to $D_0=1.6$ and $D_0=2.1$, respectively. The left and right columns correspond to $N_\mathrm{part}=2$ and $N_\mathrm{part}=3$, respectively.}
    \label{fig:PhaseDiagramsOverlap}
\end{figure}

Fig.~\ref{fig:PhaseDiagramsOverlap} shows the overlaps between the model Laughlin state and the exact $N_\mathrm{part}$-excitation ground state of a corresponding angular momentum sector for $D_0=1.6$ and $D_0=2.1$ systems with $N_\mathrm{part}=2$ and $N_\mathrm{part}=3$. The overlap is only computed if the system has LLL-like structure with $m_\mathrm{max}=2(N_\mathrm{part}-1)$. The reason is that while the second-quantized expression for the model state can be applied always, it will not have the same real-space meaning if the orbitals are fundamentally different. In Fig.~\ref{fig:PhaseDiagramsOverlap} it can be seen that not the whole parameter region is equally suitable for the Laughlin-like states (especially in the lower part in Fig.~\ref{fig:PhaseDiagramsOverlap}(d) the overlap becomes relatively low), but for every $d_0$ up to $\sim  0.3$ one can find a suitable confining potential which yields relatively high overlap. Note that the higher is the confining potential, the more the system resembles the quasi-1D minimal 6-site system.
 
The LLL-like structure as defined here is not a strict condition. For example $d_0=0.3$, $V_0=10$ for $D_0=1.6$, studied in Section \ref{sec:output} is already in the white (non-LLL-like) region of Fig.~\ref{fig:PhaseDiagramsOverlap}(a), but very close to its border. While, strictly speaking, the $\ket{\phi_2^0}$ orbital has slightly smaller average radius than $\ket{\phi_1^0}$, these states are adiabatically connected to the states of a nearby parameter region (say $d_0=0.28$, $V_0=10$) which have a LLL-like structure. However, one should not stray too far from the LLL-like region to avoid obtaining high overlaps spuriously without connection with Laughlin-like physics (after all, the model state is a state of only two particles in three orbitals).

\begin{figure*}
    \centering
    \includegraphics[width=\linewidth]{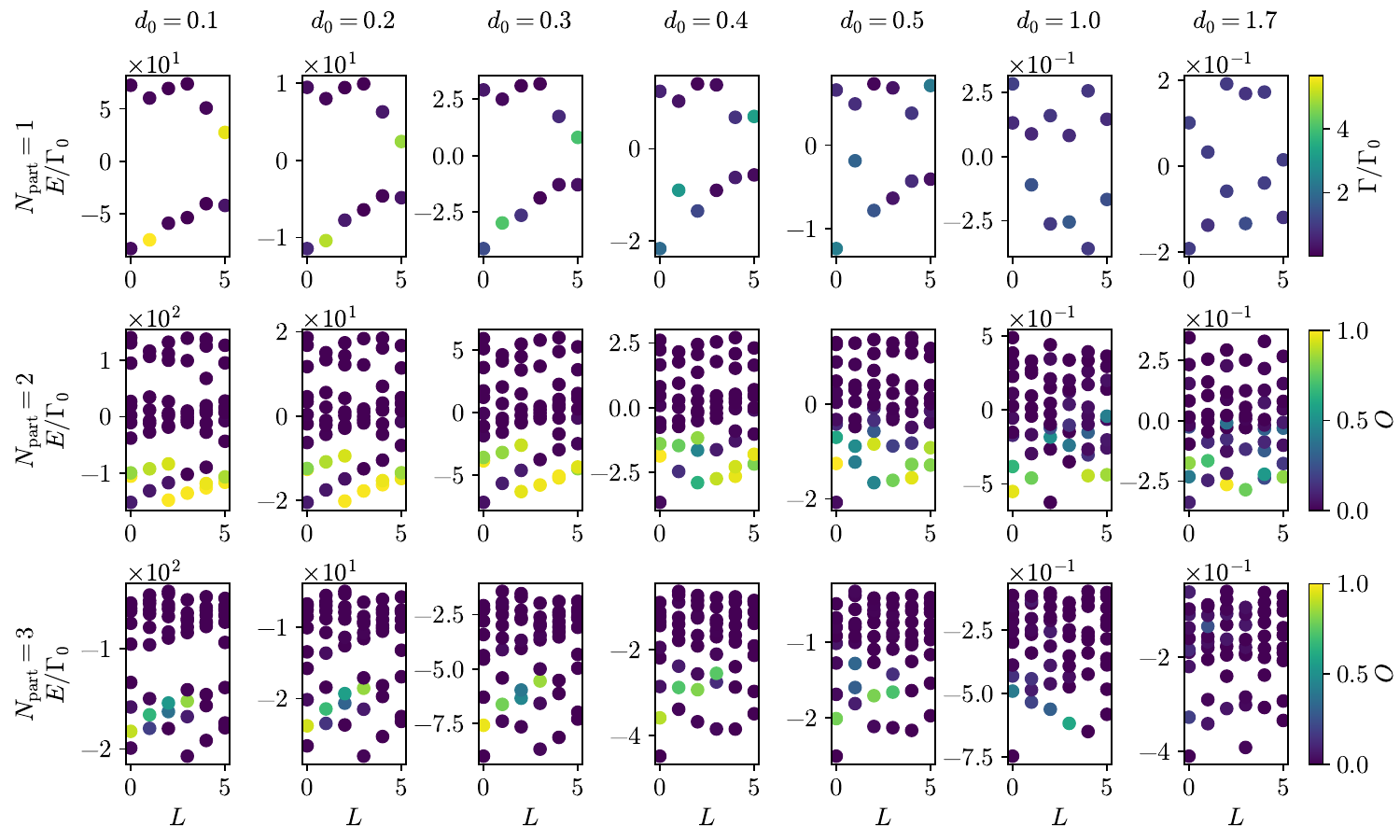}
    \caption{Energy spectra of the six-site $D_0=0.6$ system. The upper, middle and lower rows correspond to $N_\mathrm{part}=1$, $N_\mathrm{part}=2$ and $N_\mathrm{part}=3$, respectively, while the columns correspond to different values of $d_0$.  The color denotes the decay rate in the upper row and overlap with model states in the other rows. }
    \label{fig:MinimalSpectra}
\end{figure*}

As for the $D_0=0.6$ system itself, in Fig.~\ref{fig:MinimalSpectra} we plot its energy spectra with $N_\mathrm{part}=1$, $N_\mathrm{part}=2$ and $N_\mathrm{part}=3$ for various $d_0$. Here, the LLL-like structure defined in terms of radius does not apply, as each single-particle orbital has the same radius. Instead of the eigenstates, we look at the energy spectrum, which at low $d_0$ has structure similar to the LLL in a parabolic potential - in the lower band, the energy grows with $L$ (except from the $L=5$, orbital whose energy is slightly less than for $L=4$). This structure gets distorted as one increases the lattice constant. First, the $L=1$ orbital moves higher in energy, but ultimately, at high enough $d_0$, the whole ``energy growing with $L$'' structure disappears. 

This has an effect on the stability of Laughlin-like states. In the second row of Fig.~\ref{fig:MinimalSpectra} we plot the two-particle energy spectra with color denoting the overlap with model states. Similarly to Fig.~\ref{fig:Overlap}(c) of the main text, for two particles at $d_0\leq 0.3$ there is a well-visible Laughlin-like branch with high overlaps with model states (yellow and green dots). As $d_0$ is increased, this branch gets distorted and ultimately destroyed. However, it should be noted that, probably out of sheer randomness, states with high overlap with some states from the Laughlin-like branch do exist.

In the case of $N_\mathrm{part}=3$ (bottom row of Fig.~\ref{fig:MinimalSpectra}), we also observe Laughlin-like states at low $d_0$, although they are buried higher in the spectrum and do not always exist for all momentum sectors. The generalized Pauli principle for three particles on six orbitals predicts one state per angular momentum sectors $L=0,1,2,3$ and no states for other sectors. Instead, at $d_0=0.1$ we can see that at $L=2$ the overlap with the model state is ``distributed'' into two states. As for two particles, the Laughlin-like spectrum is further distorted when one increases $d_0$.

\section{Laguerre-Gauss modes at small $w_0$}\label{app:LG}

We check the validity of \eqref{eq:gaussian_mode} for $w_0\sim \lambda_A$ following \cite{manzoni2018optimization}. The idea is to modify the Laguerre-Gauss modes in a way that neglects the evanescent components, and then compare the power transmitted in the $z$ direction with \eqref{eq:gaussian_mode}. The calculation  relies on the angular spectrum decomposition
\begin{equation}
  \mathbf{E}(\mathbf{r}) = \iint \mathrm{d} p \mathrm{d} q \mathbf{A}(p,q)\exp(ik_A (px+qy+mz))
  \label{eq:ASR}
\end{equation}
with $\mathbf{A}$ being the angular spectrum component, and $p$, $q$, $m$ being rescaled $x$, $y$ and $z$ momentum coordinates, respectively (with the length of momentum vector fixed to be $k_A$). If $p^2+q^2\leq 1$, the wave is propagating and $m=\sqrt{1-p^2-q^2}$, while for $p^2+q^2 > 1$ (evanescent component) $m=i\sqrt{p^2-q^2-1}$. The $A_z$ is related to the $A_x$, $A_y$ by Maxwell's equations.

The idea is to consider a mode which has the same evanescent components as a LG mode, but has no evanescent component. Therefore, we Fourier-transform \eqref{eq:gaussian_mode} to get $A_x$ and $A_y$, set them to zero for $p^2+q^2 > 1$ and do the inverse Fourier transform to get the real-space expression. Additionally, we note that \eqref{eq:gaussian_mode} is defined in the paraxial approximation and thus it is not an exact solution to Maxwell equations. The true solution has a nonzero $z$ component, which we can determine from the Maxwell's equations. The final result for a generic polarization $\alpha \hat{\mathbf{x}}+\beta \hat{\mathbf{y}}$ is
\begin{multline}
    E_x(r, \phi)|_{z=0}= \alpha \frac{E_0}{2 f^2} \exp\left(-il\phi\right)   \sqrt{2}^{|l|}\\
\int_{0}^{1} \mathrm{d}b       b  \exp\left(-\frac{b^2}{4f^2} \right)
\left( \frac{b}{2f}\right)^{|l|} J_{|l|}(k_Abr),
\label{eq:FocusedLGX}
\end{multline}
\begin{multline}
    E_y(r, \phi)|_{z=0}= \beta \frac{E_0}{2 f^2} \exp\left(-il\phi\right)   \sqrt{2}^{|l|}\\
\int_{0}^{1} \mathrm{d}b       b  \exp\left(-\frac{b^2}{4f^2} \right)
\left( \frac{b}{2f}\right)^{|l|} J_{|l|}(k_Abr),
\label{eq:FocusedLGY}
\end{multline}
\begin{multline}
    E_z(r, \phi)|_{z=0}= 
    i(-1)^{l\Theta_H(-l) }\frac{E_0}{4 f^2}\sqrt{2}^{|l|}\\
    \left[
    \left(\alpha-i\beta\right) \exp\left(-i(l-1)\phi\right)\vphantom{\frac12}\right.\\
\int_{0}^{1} \mathrm{d}b      \frac{b^2}{\sqrt{1-b^2}}  \exp\left(-\frac{b^2}{4f^2} \right)
\left( \frac{b}{2f}\right)^{|l|}   J_{l-1}(k_Abr)\\
   -\left(\alpha+i\beta\right)\exp\left(-i(l+1)\phi\right)\\
 \left.\int_{0}^{1} \mathrm{d}b      \frac{b^2}{\sqrt{1-b^2}}  \exp\left(-\frac{b^2}{4f^2} \right)
\left( \frac{b}{2f}\right)^{|l|}   J_{l+1}(k_Abr)
\right],
\label{eq:FocusedLGZ}
\end{multline}
where $f=\frac{1}{k_A w_0}$, and $\Theta_H$ is the Heaviside function with $\Theta_H(0)=0$ (i.e. there is a sign flip in $E_z$ for odd negative $l$).

As a single number which quantifies the quality of approximation, we choose the normalization, defined in a way proportional to the power transferred in the $z$ direction, i.e. to the integral of the $z$ component of the Poynting vector $S^z$, $\iint_{z=\mathrm{const}} \mathrm{d}x  \mathrm{d}y S^z$,
\begin{equation}
F=\left(\frac{2\pi}{k} \right)^2     \iint \mathrm{d}p \mathrm{d} q \left(|\mathbf{A}\cdot \hat{\mathbf{e}}^1_{\mathbf{k}}|^2+|\mathbf{A}\cdot \hat{\mathbf{e}}^2_{\mathbf{k}}|^2\right)m,
\end{equation}
where $\hat{\mathbf{e}}^1_{\mathbf{k}}$, $\hat{\mathbf{e}}^2_{\mathbf{k}}$ are the two polarization vectors orthogonal to $\mathbf{k}$. This is consistent with the normalization $F=\iint_{z=0}\mathrm{d}x \mathrm{d} y |E_x|^2+|E_y|^2  $ of the Gaussian beam \eqref{eq:gaussian_mode}, as in that case $\mathbf{k}$ is parallel to $z$ direction. The constant $E_0$ in \eqref{eq:gaussian_mode} is chosen so that for a Gaussian beam $F=1$.

The general expression for the normalization of a circularly polarized mode is 
\begin{multline}
     F=2^{|l|-2}  \frac{\pi  E_0^2w_0^2}{f^2} \int_{0}^{\pi/2} \mathrm{d}  \theta \sin \theta \exp\left(-\frac{\sin^2 \theta}{2f^2} \right)\\
\left( \frac{\sin \theta}{2f}\right)^{2|l|} (\cos^2 \theta+1).
\label{eq:FocusedLGNorm}
\end{multline}
Using the fact that $E_0=\sqrt{\frac{2}{|l|! \pi w_0^2}}$, the coefficient in front becomes $2^{|l|-2} \frac{\pi  E_0^2w_0^2}{f^2} = \frac{2^{|l|-1}}{|l|!f^2}$.

\begin{figure}
    \centering
    \includegraphics[width=\linewidth]{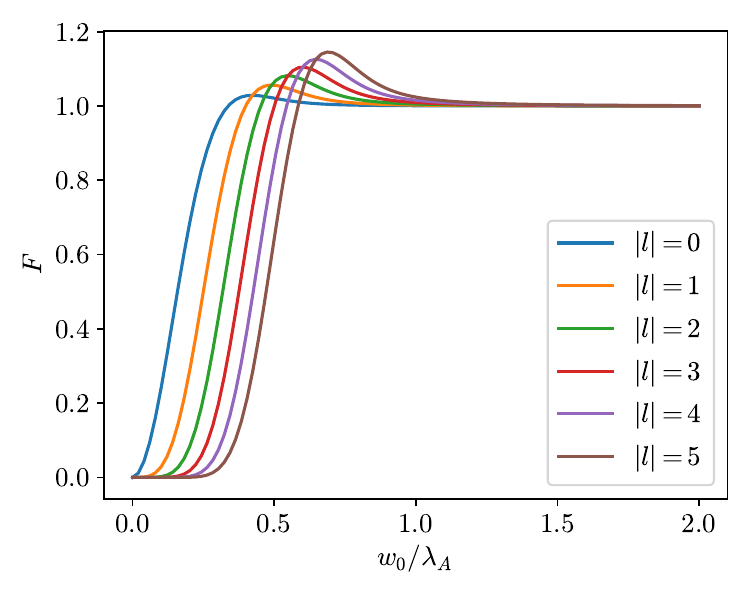}
    \caption{Normalization factors \eqref{eq:FocusedLGNorm} of the modified Laguerre-Gauss modes \eqref{eq:FocusedLGX}-\eqref{eq:FocusedLGZ} as a function of $\frac{w_0}{\lambda_A}=\frac{1}{2\pi f}$. }
    \label{fig:normalization}
\end{figure}

The solutions for $|l|=0,1,2$ are
\begin{equation}
    F_{|l|=0}=\frac{1}{2}  \left( 1  +\sqrt{2} \left(-  f     +  \frac{1}{f} \right)D_+\left(\frac{1}{f\sqrt{2}}\right)\right)
\end{equation}
\begin{equation}
    F_{|l|=1}= 
\frac{3}{4}-\frac{1}{4f^2}+\frac{\sqrt{2}}{4f^3}D_+\left(\frac{1}{f\sqrt{2}}\right)-\frac{3\sqrt{2}}{4} f D_+\left(\frac{1}{f\sqrt{2}}\right)
\end{equation}
\begin{multline}
    F_{|l|=2}=    \frac{\sqrt{2}}{4} \left[
-\frac{1}{4\sqrt{2}f^4} -\frac{2}{4\sqrt{2} f^2}+\frac{15}{4\sqrt{2}} \right.  \\-\frac{15f}{4} D_+\left(\frac{1}{f\sqrt{2}} \right)
-\frac{3}{4f}D_{+} \left(\frac{1}{f\sqrt{2}}  \right) \\ \left.
+\frac{1}{4f^3}D_{+}\left(\frac{1}{f\sqrt{2}} \right) +
\frac{1}{4f^5} D_+\left(\frac{1}{f\sqrt{2}} \right)
\right],
\end{multline}
where $D_{+}(x)=e^{-x^2}\int_0^{x}e^{t^2} \mathrm{d}t$ is the Dawson function. As $|l|$ increases, the expressions become more and more complicated, but they still involve powers of $f$ and Dawson functions, and they can be calculated using SymPy. The normalization as a function of $w_0$ is shown in Fig.~\ref{fig:normalization}. It can be seen that the deviation of the asymptotic value (unity) grows with $|l|$. Nevertheless, it seems that at $w_0=\lambda_A$, we have  $F\approx 1$ for all $|l|<6$, suggesting that we can still approximate the modes with \eqref{eq:gaussian_mode}.



\section{More results on $D_0=1.6$ system}\label{app:bigger}

In Sec.~\ref{sec:driving} we have shown that it is possible to realize a Laughlin-like state in a driven $D_0=1.6$ system and that its presence is visible from rescaled two-photon intensities and signatures. Here, we provide more results to show how this situation changes if one increases the lattice constant. 

\begin{figure}
    \centering
    \includegraphics[width=\linewidth]{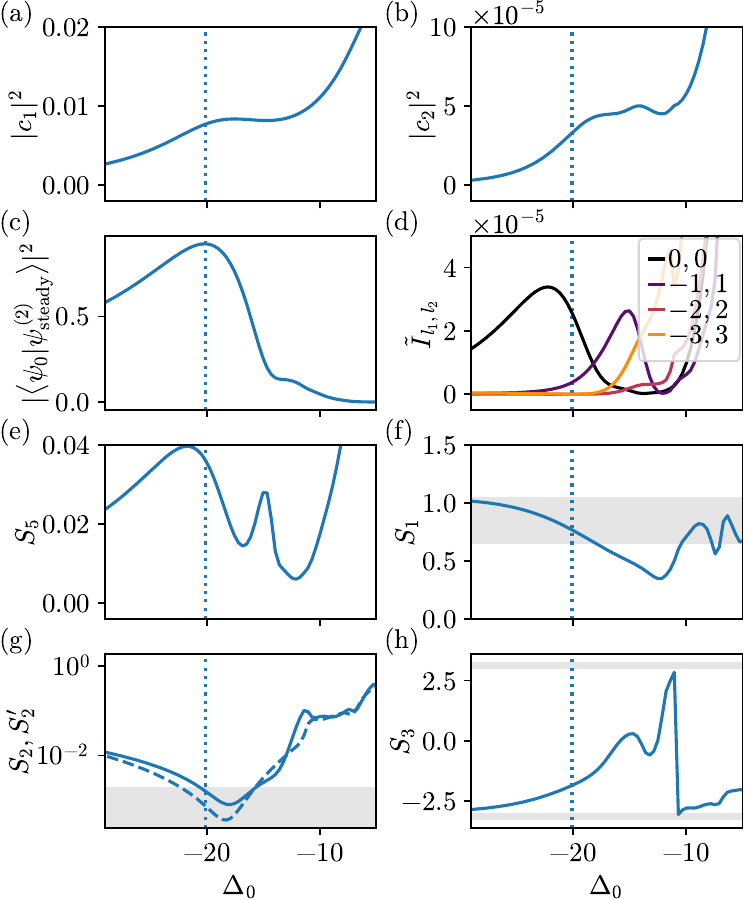}
    \caption{Driving the $D_0=1.6$, $d_0=0.2$, $V_\mathrm{harm0}=3$ system with spatially uniform Rabi frequency $\Omega=0.1\Gamma_0$. (a), (b) the square modulus of the 1- and 2-excitation coefficients $c_1$, $c_2$, respectively. (c) Squared overlap between the two-excitation part of steady state and the static ground state. (d) Rescaled two-photon intensity $\tilde{I}_{l_1,l_2}$. (e)-(h) Signatures $S_5$, $S_1$, $S_2$ (with $S_2'$) and $S_3$, respectively (Eqs. \eqref{eq:S1}-\eqref{eq:S3} and \eqref{eq:S5}). Solid and dashed lines in (g) denote $S_2$ and $S_2'$, respectively. The dotted vertical lines in all the subplots denote the position of the maximum of overlap.}
    \label{fig:DrivingSignaturesBigger0.2}
\end{figure}

\begin{figure}
    \centering
    \includegraphics[width=\linewidth]{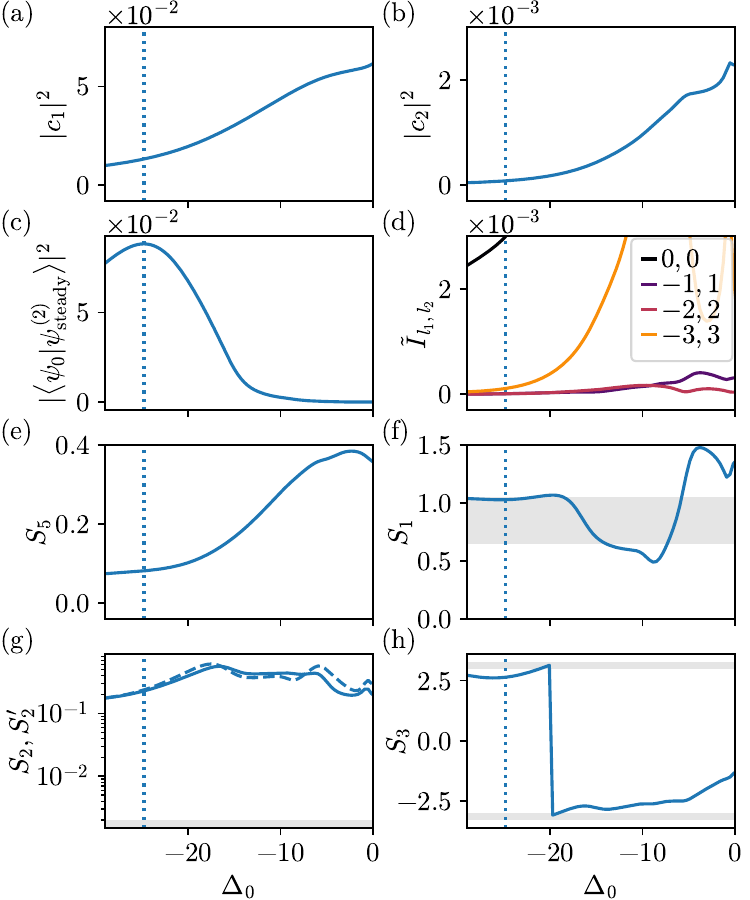}
    \caption{Driving the $D_0=1.6$, $d_0=0.3$, $V_\mathrm{harm0}=10$ system with spatially uniform Rabi frequency $\Omega=0.1\Gamma_0$. (a), (b) the square modulus of the 1- and 2-excitation coefficients $c_1$, $c_2$, respectively. (c) Squared overlap between the two-excitation part of steady state and the static two-excitation ground state. (d) Rescaled two-photon intensity $\tilde{I}_{l_1,l_2}$. (e)-(h) Signatures $S_5$, $S_1$, $S_2$ (with $S_2'$) and $S_3$, respectively (Eqs. \eqref{eq:S1}-\eqref{eq:S3} and \eqref{eq:S5}). Solid and dashed lines in (g) denote $S_2$ and $S_2'$, respectively. The dotted vertical lines in all the subplots denote the position of the maximum of overlap.}
    \label{fig:DrivingSignaturesBigger0.3}
\end{figure}

Figs.~\ref{fig:DrivingSignaturesBigger0.2} and \ref{fig:DrivingSignaturesBigger0.3} are analogous to Fig.~\ref{fig:DrivingSignaturesBigger} of the main text, with the difference that now $d_0=0.2$ and $d_0=0.3$, respectively. In the case of $d_0=0.2$, the maximum squared overlap between the two-particle sector of the steady state and the static two-excitation $L=2$ ground state is  $\sim 0.92$ (Fig.~\ref{fig:DrivingSignaturesBigger0.2}(c)). The similarity to the Laughlin state can be detected using rescaled two-photon amplitudes: at the maximum overlap (and at the maximum of $S_5$, which is at a slightly smaller value of detuning) the $\tilde{I}_{0,0}$ and $\tilde{I}_{-1,1}$ components dominate after rescaling (Fig.~\ref{fig:DrivingSignaturesBigger0.2}(d)). Also, the signatures $S_1$, $S_2$, $S_2'$ lie within the gray regions or close to them (Fig.~\ref{fig:DrivingSignaturesBigger0.2}(e),(f)). In the case of $S_3$, one can see a sizable departure from the gray region (Fig.~\ref{fig:DrivingSignaturesBigger0.2}(g)), which signifies that the state is not completely similar to the Laughlin state.

In the case of $d_0=0.3$, the two-particle sector of the steady state is no longer Laughlin-like, as indicated by small overlap with static $L=2$ ground state (Fig.~\ref{fig:DrivingSignaturesBigger0.3}(c)). The maximum of $S_5$ no longer detects maximum overlap. However, both in the maximum of overlap and in the maximum of $S_5$, the observables point at the absence of the Laughlin state. In the rescaled intensities, $\tilde{I}_{-3,3}$ dominates over $\tilde{I}_{-1,1}$ in the whole plotted range of detunings. Similarly, in the whole range the signatures $S_2$, $S_2'$ are far from the gray region. 

\bibliography{tfb}
\end{document}